\documentclass[aip,pof,reprint,amsmath,amssymb]{revtex4-1}
\pdfoutput=1

\usepackage{graphicx}

\usepackage{amsmath}

\usepackage{bm}

\usepackage{ifpdf}

\usepackage{caption}
\usepackage{subcaption}

\usepackage{color}
\usepackage[usenames,dvipsnames]{xcolor}

\DeclareMathOperator*{\define}{\equiv}

\newcommand{\eq}[1]{ Eq.\ (\ref{#1})}

\newcommand{\figwidth}{0.7}

\newcommand{\ie}{\textit{i.e.,}}

\newcommand*{\xline}[1][1.0em]{\rule[0.4ex]{#1}{0.5pt}}
\newcommand*{\xlinethick}[1][1.0em]{\rule[0.4ex]{#1}{1.5pt}}

\newcommand*{\xlineshort}[1][1.0em]{\rule[0.4ex]{3.5pt}{0.5pt}}

\newcommand*{\xdash}[1][1.0em]{\rule[0.01ex]{2.5pt}{0.5pt} \rule[0.01ex]{0.5pt}{0.5pt}}
\newcommand*{\xdashthick}[1][1.0em]{\rule[0.5ex]{2.5pt}{1.5pt} \; \rule[0.5ex]{2.5pt}{1.5pt}}
\newcommand*{\xdashvthick}[1][1.0em]{\rule[0.4ex]{4.5pt}{2.5pt} \rule[0.4ex]{4.5pt}{2.5pt}}

\definecolor{lllgrey}{rgb}{0.9,0.9,0.9}
\definecolor{llgrey}{rgb}{0.8,0.8,0.8}
\definecolor{lgrey}{rgb}{0.6,0.6,0.6}
\definecolor{dlgrey}{rgb}{0.4,0.4,0.4}
\definecolor{ddlgrey}{rgb}{0.2,0.2,0.2}
\definecolor{cm_orange}{rgb}{0.9950,0.8090,0.5000}
\definecolor{cm_red}{rgb}{0.9520,0.4180,0.2580}
\definecolor{cm_blue}{rgb}{0.3680,0.5750,0.7660}

\definecolor{fft_darkblue}{rgb}{0.19215686619281769, 0.21176470816135406, 0.58431375026702881}
\definecolor{fft_lightblue}{rgb}{0.56485968018118948, 0.7663975529283894, 0.86758939636894861}
\definecolor{fft_yellow}{rgb}{0.99992310649750094, 0.99761630142252977, 0.74540562818527956}
\definecolor{fft_lightred}{rgb}{0.97347174068223274, 0.54740483716834987, 0.31810842222408564}
\definecolor{fft_darkred}{rgb}{0.64705884456634521, 0.0, 0.14901961386203766}

\begin{document}

\title{A Molecular Dynamics Simulation of the Turbulent Couette Minimal Flow Unit}

\author{E.~R.~Smith}
\affiliation{Department of Mechanical Engineering, Imperial College London, 
Exhibition Road, South Kensington, London SW7 2AZ, United Kingdom} 

\date{\today}

\begin{abstract}

A molecular dynamics (MD) simulation of planar Couette flow is presented for the minimal channel in which turbulence structures can be sustained. 
Evolution over a single breakdown and regeneration cycle is compared to computational fluid dynamics (CFD) simulations. 
Qualitative similar structures are observed and turbulent statistics show excellent quantitative agreement.
The molecular scale law of the wall is presented in which stick-slip molecular wall-fluid interactions replace the no-slip conditions.
The impact of grid resolution is explored and the observed structures are seen to be dependant on averaging time and length scales. 
The kinetic energy spectra show a range of scales are present in the molecular system and that spectral content is dependent on the grid resolution employed.
The subgrid velocity of the molecules is compared to spatial averaged velocity using joint probability density functions.
Molecular trajectories, diffusions and Lagrangian statistics are presented.
The importance of sub-grid scales, relevance of the Kolmogorov lengthscale and implications of molecular turbulence are discussed.

\end{abstract}

\pacs{}

\maketitle

\section{Introduction}

Molecular and turbulent motions are traditionally thought to exist at very different scales, separated by three or more orders of magnitude \citep{Pope}.
As a result, turbulence at the molecular scale has not been the focus of significant study \citep{Muriel}, in large part due to the prohibitive molecular system sizes required. 
However, as will be shown in this work, through careful choice of viscosity and geometry a turbulent molecular simulation is possible with current computing resources.

The traditional simulation tool for turbulent flow is continuum computational fluid dynamics (CFD) \citep{Hirsch}.
CFD models the flow of fluids away from hydrodynamic equilibrium, using the Navier-Stokes equation. 
The Navier-Stokes equation is based on Newton's law, with additional assumptions, namely: 
a Newtonian framework in an inertial reference frame, 
thermodynamic equilibrium (or quasi-equilibrium), fluid isotropy, 
stress tensor symmetry and sometimes incompressibility \citep{Gad-El-Hak_06}. 
Only with the advent of computers have general numerical solutions of the Navier-Stokes equations been possible. 
With these numerical solutions, complex and apparently random behavior is the norm, a phenomenon known in fluid mechanics as turbulence. 
Turbulence is a feature of almost every case of engineering interest. 
Greater understanding of this phenomenon would facilitate improvements in predictive capabilities and advances in engineering design. 

\begin{figure}
\includegraphics[width=\figwidth\textwidth]{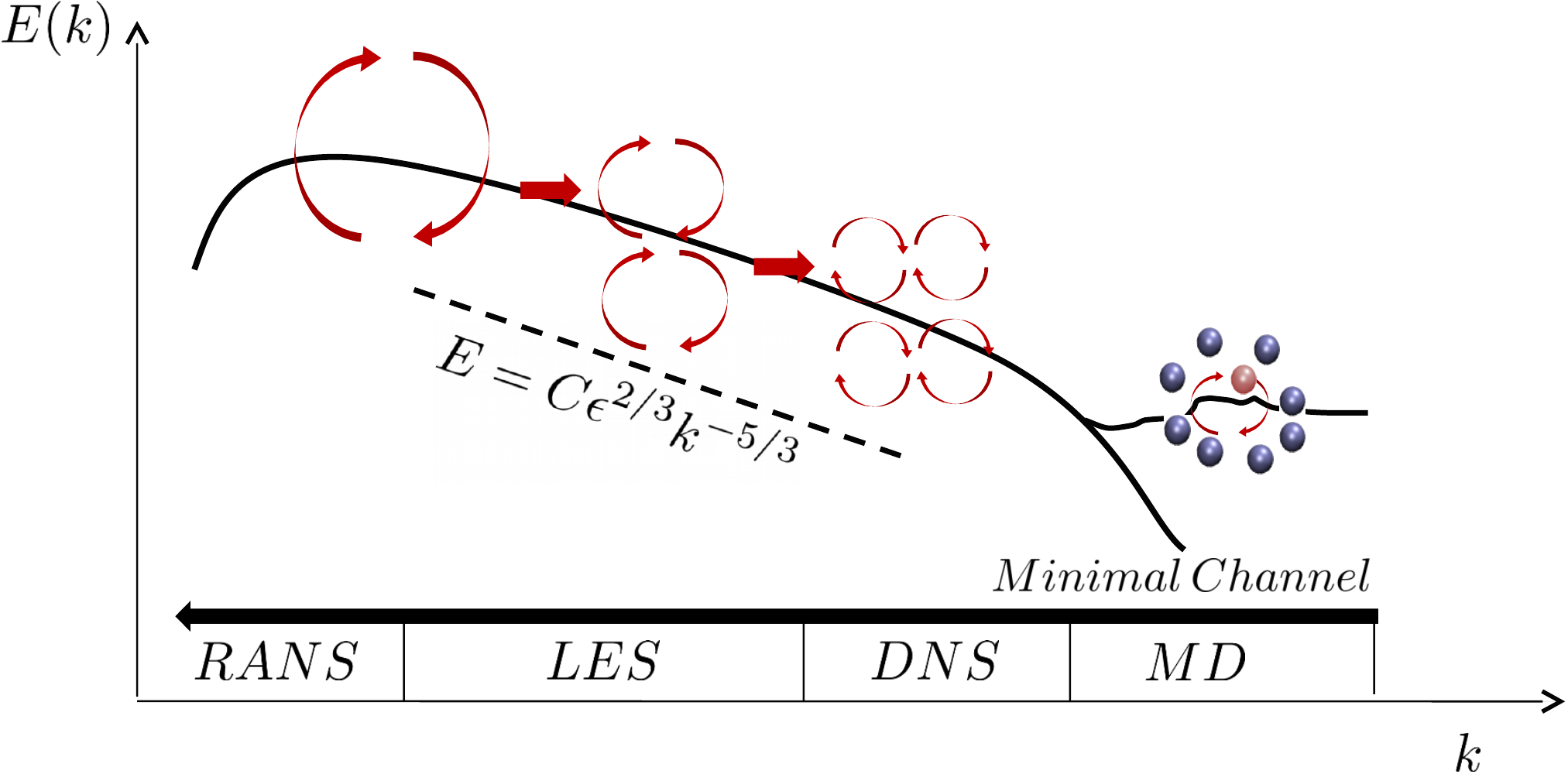}
\caption{Schematic of the turbulent energy cascade and validity of modeling \citep{Hirsch}, including Molecular dynamics (MD). 
Moving left to right, each successive model contains more physical scales of fluid motion but is limited by computational cost to simulation of a smaller maximum lengthscale.}  
\label{cascade_schematic}
\end{figure}

Non-equilibrium molecular dynamics (NEMD) \citep{Evans_Morris} is an alternative approach to fluid modeling which requires only Newton's laws and a form of inter-molecular potential. 
As a more fundamental model, molecular dynamics (MD) captures a much greater range of phenomena with no additional assumptions.
Examples include: 
arbitrarily strong shock waves \citep{Hoover_1979}, 
bubble nucleation, phases change and co-existence \citep{Okumura_03}, moving three-phase contact lines \citep{Thompson_et_al93}  
exact energy conservation \citep{Tuckermann_book}, 
visco-elasticity or memory effects \citep{Evans_Morris1986} and detailed solid-liquid interface \citep{Hoover,Evans_Morris}. 
The price for this generality is in computational overhead, limiting accessible systems to the microscale. 
Despite this, many complex fluid dynamical phenomena have been reproduced at the molecular scale, with excellent agreement to the continuum.
These include Poiseuille and Couette flow \citep{Evans_Morris}, vortex shedding on a cylinder \citep{Rapaport_87} or plate \citep{Cui_Evans,Meiburg}, 
Taylor-Couette rolls \citep{Rapaport_98,Rapaport_2000, Rapaport_2014}, the Rayleigh-Taylor 
instability \citep{Rapaport_88,Kadau20042004}, two dimensional fluid mixing \citep{Dzwinel2000} as well as turbulence like vortices due to 
shock waves \citep{Root_et_al}. 
The fundamental nature of MD means it is uniquely placed to provide insight into the mechanisms of turbulent transport. 
Figure \ref{cascade_schematic} shows a schematic of the spectral energy cascade, believed to be a central feature of turbulent flow \citep{Pope}.
Large scales eddying motions break down into smaller scales until at some minimum length scale the coherent motions dissipate to heat as a consequence of viscosity.
Energy is conserved in an MD simulation and the viscous heating is simply the change of coherent velocity to non-coherent molecular fluctuations.
Molecular simulation requires no grid and the minimum flow scale is dictated by the underlying configuration of the molecules themselves. 
As a result, molecular modeling of turbulence has the potential to provide a complete picture of the energy cascade and link thermodynamic concepts such as pressure and temperature along with viscosity to hydrodynamic fluctuations.
The range of application of various fluid modeling methodologies are included on Figure \ref{cascade_schematic}.
Reynolds Averaged Navier Stokes (RANS) is the most commonly used industrial solution to fluid problems and models only the largest scales.
Molecular simulation can be viewed as a refinement to direct numerical simulation (DNS), where sub-dissipative scales are modelled with no filtering of thermodynamic `noise'.
This is analogous to Large Eddy Simulation (LES) which filters the smaller turbulent motions of a DNS, replacing them with a closure model to approximate the details.
From this viewpoint, viscosity and pressure in a continuum solution are simply a closure model to replace the filtered molecular detail.
There is therefore a clear benefit of MD to industrial fluid mechanics, especially in the rapidly developing field of nano and micro fluidics.
As with DNS simulation, MD can be used to refine and improve the underlying models which are essential to industrial CFD simulation.

In this work, we aim to explore the minimum scale at which turbulence breaks down into viscous heat.
We reduce the system size until it contains only the simplest turbulent flow -- known as the minimal flow-unit in the literature \citep{jimenez_moin,Hamilton}.
The smallest turbulent flow-unit is found in planar Couette flow, where a viscous fluid is driven by two infinite plates sliding in opposite directions.
This simulated case is an example of wall-bounded turbulent shear flow and, due to the low Reynolds number, only models flow in the near-wall region. 
The flow is not fully turbulent and as a result, the observed flow structures are well organized and a repeating cycle of streak `busting' and reformation is observed.
In a recent summary, \citet{Jimenez13} concludes that available evidence strongly suggests minimal flow units are representative of dynamics in real large scale turbulent flow.
Experimental evidence also shows that the details of near-wall bursting are identical for a range of Reynolds numbers \citep{Viswanath}.
The minimal channel therefore possess the main attributes of turbulent flow, namely that it is non-laminar, three dimensional, stochastic \citep{jimenez_moin} and shows significant and irregular variations in both space and time \citep{Pope}.
More importantly, due to its minimal size, it is computationally tractable using molecular dynamics.

The first part of the work introduces the modeling methodologies of computational fluid dynamics and molecular dynamics. 
A parameter study is then presented to determine the optimal density and temperature to enable a computationally feasible turbulence simulation. 
Analysis of the flow is performed by identifying identical time evolving turbulent structures in both the molecular and continuum systems. 
The streak breakdown and vortex regeneration of the minimum channel flow are demonstrated in the molecular system.
The phase space evolution is compared for the two systems showing the molecular system is behaving in a similar manner to the continuum and that the dynamics are non-laminar.
This is then followed by a range of standard turbulence statistical analysis. 
The existence of turbulence at the molecular scale is demonstrated through the agreement between the molecular and continuum solutions. 
Next, a revised law of the wall is presented including the molecular stacking region which is unique to molecular simulation.
The evolution of various forms of energy is plotted and the interchange between hydrodynamic and thermodynamic energies is discussed.
The impact of different grid-resolutions on the kinetic energy spectrum is explored.
Joint probability density functions and Lagrangian statistics are employed to gain insight into the sub-grid motions of the molecules.
This work ends with a discussion of the minimum scales of turbulence and the importance of MD as a tool in gain insight into fluid dynamical flows.

\section{Methodology}

In this section, the continuum model is presented first to outline how the minimal Couette channel is commonly simulated.
An initial condition is obtained, which is used in both the continuum and molecular study.
Next, the molecular dynamics model is presented and a method to obtain the required viscosity (and Reynolds number) through selection of temperature and density is discussed.
Finally, the setup and geometry of the molecular minimal channel case is outlined.

\subsection{Computational Fluid Dynamics}

Computational fluid dynamics (CFD) is the numerical simulation of fluid motion with dynamics obtained from Newton's law. The equations for an Eulerian control volume $V$,
\begin{align}
\frac{\partial }{\partial t}  \int_V  \rho \boldsymbol{u} dV \! = -
\oint_S \left[ \rho \boldsymbol{u} \boldsymbol{u} + \boldsymbol{\Pi} \right]
\cdot d\textbf{S}  + \textbf{F}_{\textnormal{body}}.
\label{BofMEqn2}
\end{align}
where $\rho$ is the fluid density,  $\boldsymbol{u}$ is the fluid velocity, $\boldsymbol{\Pi}$ the pressure tensor and $\textbf{F}_{\textnormal{body}}$ represents external body forces. 
This can be written at a point in space by applying the divergence theorem, assuming a continuum, taking the limit of zero volume and neglected body forces,
\begin{align}
\frac{\partial }{\partial t}   \rho \boldsymbol{u} \! = -
\boldsymbol{\nabla} \cdot \left[ \rho \boldsymbol{u} \boldsymbol{u} + \boldsymbol{\Pi} \right].
\end{align}
By assuming a stress strain closure relation (with stress isotropy, symmetry and thermodynamic equilibrium), the pressure tensor, $\boldsymbol{\Pi}$, can be approximated in terms of scalar pressure $P$ and velocity, 
\begin{align}
\boldsymbol{\Pi} =  P\boldsymbol{I} - \mu\left[\boldsymbol{\nabla} \boldsymbol{u} + (\boldsymbol{\nabla} \boldsymbol{u})^T - \left(\boldsymbol{\nabla} \cdot \boldsymbol{u} \right) \boldsymbol{I}\right] - \lambda \left(\boldsymbol{\nabla} \cdot \boldsymbol{u}\right) \boldsymbol{I}.
\label{STEqn}
\end{align}
using Stokes hypothesis $ \lambda + 2/3\mu = 0$ and fluid incompressibility, $\boldsymbol{\nabla} \cdot  \boldsymbol{u} = 0$, the Navier-Stokes equations are obtained,
\begin{align} 
\frac{\partial \boldsymbol{u}}{\partial t} + 
\left( \boldsymbol{u} \cdot \boldsymbol{\nabla} \right) {\boldsymbol u}
= -  \boldsymbol{\nabla} P + \frac{1}{Re} \nabla^2 \boldsymbol{u},
\label{ANDNSEqn}
\end{align}
where the formulas in \eq{ANDNSEqn} are non-dimensionalised by the channel half height $h$, wall velocity $U_w = \pm 1$ and the kinematic viscosity $\nu_{_{\text{CFD}}}=\mu_{_{\text{CFD}}} / \rho_{_{\text{CFD}}}$ with $Re_{_{\text{CFD}}} \define h U_w / \nu_{_{\text{CFD}}}$. 
Channelflow \citep{channelflow}, a spectral solver is employed to numerically approximate the solution to the incompressible and isothermal Navier Stokes \eq{ANDNSEqn}. 
Channelflow's numerics are based on \citet{Canuto_et_al}, and solve the Navier Stokes equations with a Chebyshev-tau algorithm in the primitive variable formulation.
To simulate planar Couette flow, the $y$ boundary conditions model counter-sliding walls using the no-slip condition, $u(\pm h)=\pm U_w$.
The wall boundary conditions and incompressibility are enforced using the tau correction \citep{Canuto_et_al} with periodic boundaries in the $x$ and $z$ directions.
The velocity field is defined on $256$ Gauss-Lobatto points in $y$ with $64$ by $64$ uniform points in the $x$ and $z$ directions. Chebyshev basis functions are employed in the wall normal and Fourier components in the spanwise and streamwise directions.  
The rotational form of the non-linear advection term is used with $2/3$ dealiasing. 
Time advancement is implemented using third order semi-implicit backwards differencing. 
A Reynolds number of four hundred is chosen in line with the existing literature \citep{Hamilton,GibsonHalcrowCvitanovicJFM08,Viswanath}, supported by 
experiments \citep{Bech_et_al} which show turbulence for Reynolds number as low as $360$. 
The channel size is chosen to be $L_x \times L_y \times Lz = 1.75\pi h \times 2h \times 1.2 \pi h $ in units normalised by the channel half height \citep{Hamilton}. 
Channelflow is ideally suited for this geometry and has been used extensively to simulate the minimal flow unit \citep{GibsonHalcrowCvitanovicJFM08}.

The Reynolds number for the minimum channel is sub-critical and as a result, will not naturally transition even in the presences of significant perturbations \citep{jimenez_moin}.
As in the original work of \citet{Hamilton}, turbulent flow is obtained by initializing a high Reynolds number case with random perturbations and reducing the Reynolds number.
This process is used here, with $Re= 1000$, $700$, $500$ and finally $400$.
The CFD field at $Re=400$ was then run for $2000 h/U$ time units to ensure any artifacts of the initialization were washed out and that turbulence is sustained. 
The flow field was then saved to provide the initial velocity condition for both the MD and CFD runs. 
Channelflow was started from the same initial condition and was run for a single flow through time, $100 h/U$, with detailed files collected to allow comparison with the MD solution over the same time.
As the continuum minimal channel is driven by the wall shear, the regeneration cycle appears to occur indefinitely.
A further run over $35$ regeneration cycles ($3500 h/U$ flow though times) was then carried out to collect well resolved statistics.

Reynolds decomposition is employed in order to analyze the evolution of the flow. 
The mean velocity, $\overline{\boldsymbol{u}}$, in the channel is defined by averaging over homogeneous directions in both space and time.
For some arbitrary quantity $A$, 
\begin{align}
\overline{\mathcal{A}}(y) \define \frac{1}{\tau_{_{CFD}} L_x L_z} \int_0^{L_x} \int_0^{L_z} \int_0^{T} \mathcal{A}(x,y,z,t) dt dx dz,
\label{meanvel}
\end{align}
where $\tau_{_{CFD}}$ is the entire simulation time.
Using the definition of mean velocity from \eq{meanvel}, the velocity perturbation can be defined as $\boldsymbol{u}^{\prime} = \boldsymbol{u} - \overline{\boldsymbol{u}}$

\subsection{Molecular Dynamics}
Molecular dynamics involves the solution of Newton's law for every molecule in an N-molecular system,
\begin{align}
 m_i \boldsymbol{\ddot{r}}_i =\textbf{F}_i,
\label{Newtonslaw}
\end{align}
where $m_i$ is the mass of molecule $i$, $\boldsymbol{\ddot{r}}_i$ is its acceleration and the force it experiences, $\textbf{F}_i$, is due to the sum of all intermolecular interactions,
$\textbf{F}_i = \sum_{i \ne j} \boldsymbol{f}_{ij} =  
\sum_{i \ne j}\boldsymbol{\nabla} \varPhi_{ij}$.
Here $\varPhi_{ij}$ is the intermolecular potential function,
\begin{align}
 \varPhi(\boldsymbol{r}_{ij}) =
\begin{cases}4\epsilon \left[ \left( 
 \frac{\ell}{\boldsymbol{r}_{ij}} \right) ^{12} 
 - \left( \frac{\ell}{\boldsymbol{r}_{ij}} \right) ^{6} \right] - 4\epsilon
\left[ \left( 
 \frac{\ell}{\boldsymbol{r}_{c}} \right) ^{12} 
 - \left( \frac{\ell}{\boldsymbol{r}_{c}} \right) ^{6} \right], & r_{ij} < r_c \\ 
0, &  r_{ij} \ge r_c  
\end{cases}
\label{LJ}
\end{align}
where $\ell$ is the molecular length scale, $\epsilon$ the energy scale, $m_i$
the mass and $\boldsymbol{r}_{ij} = \boldsymbol{r}_i - \boldsymbol{r}_j$ the 
inter-molecular separation. 
The pairwise Lennard-Jones potential, \eq{LJ}, was used with a cutoff of $r_c = 2^{\frac{1}{6}}$, which is efficient while retaining much of the essential physics \citep{Rapaport,Allen_Tildesley}. 
The Verlet time integration was employed due to its excellent energy conservation properties \citep{Allen_Tildesley}.
The software used in the investigations is designed especially for the NEMD style problem presented here, highly optimized for efficient large scale parallel computing and thoroughly verified \citep{Smith_Thesis, Smith_et_al,Heyes_et_al14}. 
The setup for planar Couette flow is well established in the non-equilibrium MD literature \citep{Heyes_et_al,Bernardi10a}.
This includes tethered wall molecules \citep{Petravic_Harrowell} with temperatures controlled using the Nos\'{e} Hoover thermostat \citep{Nose,Hoover_NoseHooverthermostat}.
The equations of motion for the wall atoms are given by,
\begin{subequations}
\begin{eqnarray}
\boldsymbol{v}_{i}  &=& \frac{\overline{\boldsymbol{p}}_{i}}{m_i} +
U_\text{w} \textbf{n}_x, \\
\dot{\overline{\boldsymbol{p}}}_{i}  &=& \boldsymbol{F}_{i} + 
\boldsymbol{F}_{i_{\textnormal{teth}}} - \xi \overline{\boldsymbol{p}}_{i}, \\
\boldsymbol{F}_{i_{\textnormal{teth}}} &=& \boldsymbol{r}_{i_0}  \left( 4 k_{4}
r_{i_0}^{2}+6 k_{6} r_{i_0}^{4} \right), \\
\dot{\boldsymbol{r}}_{i_{0}}  &=& U_\text{w} \textbf{n}_x , \\
\dot{\xi} &=& \frac{1}{Q_\xi} \left[  \displaystyle\sum_{n=1}^{N} 
\frac{\overline{\boldsymbol{p}}_{n} \cdot \overline{\boldsymbol{p}}_{n}}{m_n} -
3T_0 \right], \label{NH_verify} 
\end{eqnarray}
\end{subequations}
where $\boldsymbol{v}_{i} \define \dot{\boldsymbol{r}}_{i}$ is the molecular velocity; $\overline{\boldsymbol{p}}_i/m_i$ is the peculiar velocity in a reference frame moving at the wall speed $U_w$ with $\textbf{n}_x$ the unit vector in the $x$ direction; $T_0$ is the wall thermostat setpoint; $\boldsymbol{r}_{i_0} = \boldsymbol{r}_i - \boldsymbol{r}_0$ the departure of a wall atom from its tethering site $\boldsymbol{r}_0$ and $Q_\xi$ is the heat bath mass for the Nos\'{e} Hoover thermostat.
The strength of wall tethering is based on the work of \citet{Petravic_Harrowell} with coefficients of $k_4 = 5 \times 10^3$ and $k_6 = 5 \times 10^6$.
Only the walls are thermostatted in order to minimize unphysical effects in the dynamics of the fluid \citep{Bernardi10b}. 

The Irving and Kirkwood method \citep{Irving_Kirkwood}, together with the fluid mechanics concept of a control volume can be used to express a molecular system in the same form as the continuum system.
The mass in a control volume is,
\begin{align}
\int_V \rho  dV = \bigg\langle \displaystyle\sum_{i=1}^{N}   m_i  \vartheta_i \bigg\rangle, 
\label{mass}
\end{align}
where $\vartheta_i$ is defined in terms of a combination of Heaviside functionals so that $\vartheta_i = 1$ when a molecule $i$ is inside a control volume and $\vartheta_i=0$ when $i$ is outside \citep{Smith_et_al}. 
The angular brackets denote an average.
The momentum is,
\begin{align}
\int_V \rho \boldsymbol{u} dV = \bigg\langle \displaystyle\sum_{i=1}^{N}   m_i \boldsymbol{v}_i \vartheta_i \bigg\rangle. 
\label{momentum}
\end{align}
With $\boldsymbol{v}_i \define \dot{\boldsymbol{r}}$ denoting the laboratory velocity.
From the momentum and mass, the so called streaming velocity can be defined as,
\begin{align}
 \boldsymbol{u} \define \bigg\langle \displaystyle\sum_{i=1}^{N}   m_i \boldsymbol{v}_i \vartheta_i \bigg\rangle \bigg/ \bigg\langle \displaystyle\sum_{i=1}^{N}   m_i \vartheta_i \bigg\rangle 
\label{MD_momentum}
\end{align}
The streaming velocity is assumed to be equivalent to the hydrodynamic fluid velocity (continuum Eqs. \ref{BofMEqn2}-\ref{ANDNSEqn}).
By subtracting this from the velocity of a molecule, a definition of kinetic temperature can be obtained,
\begin{align}
T \define \bigg\langle \frac{1}{3N} \displaystyle\sum_{i=1}^{N}   m_i \left|\frac{\boldsymbol{p}_i}{m_i}\right|^2  \bigg\rangle 
\label{MD_temperature}
\end{align}
where $\boldsymbol{p}_i/m_i = \dot{\boldsymbol{r}}_i - \boldsymbol{u}$ is called the peculiar velocity.
Definition of a streaming and peculiar velocity splits the laboratory velocity (and associated kinetic energy) of the molecule into a thermal and hydrodynamic component.

The pressure tensor in a molecular system can be calculated by integrating the \citet{Irving_Kirkwood} equations over a cubic control volume, \citep{Smith_et_al},
\begin{align}
\frac{d}{dt} \displaystyle\sum_{i=1}^{N} m_i \boldsymbol{v}_i \vartheta_i = \oint_S \left[ - \rho \boldsymbol{u} \boldsymbol{u} + \boldsymbol{\Pi} \right] \cdot d\textbf{S}
\label{momentum_equation}
\end{align}
The molecular form of the pressure tensor, $\boldsymbol{\Pi}$, is then given by,
\begin{align}
\oint_S \boldsymbol{\Pi} \cdot d\textbf{S} = 
- \underbrace{  \bigg\langle \displaystyle\sum_{i=1}^{N} 
\frac{\boldsymbol{p}_i \boldsymbol{p}_i}{m_i }   \cdot  
d\textbf{S}_i }_{\text{Kinetic}} + 
\underbrace{\frac{1}{2} \displaystyle\sum_{i,j}^{N}
\textbf{f}_{ij} \textbf{n} \cdot  d\textbf{S}_{ij} \bigg\rangle
 }_{\text{Configurational}} 
\label{Stress}
\end{align}
The functional $d\textbf{S}_i$ in the kinetic term selects only molecules which have crossed a control volume surface during a time period $\tau_{_{MD}}$.
The kinetic part of the pressure tensor in \eq{Stress} is therefore the kinetic theory definition; namely a force due to the cumulative effect of molecules bouncing off a container surface \citep{Smith_et_al}.
The functional $d\textbf{S}_{ij}$ selects inter-molecular forces when molecules $i$ and $j$ are on opposite sides of the control volume surface (and the inter-molecular force is acting through the surface). 
The kinetic and configurational terms together are the method of planes form of pressure \citep{Todd_et_al_95} localized to the surface of a control volume.
Mathematically \eq{Stress} is expressed in a weakened form and is equivalent to the continuum control volume equation (\ref{BofMEqn2}).
Note that the term pressure and stress are used interchangeably at the molecular scale (where stress is simply negative pressure), although it is natural to speak of the kinetic terms as a pressure and the configurational part as a stress.

\begin{figure}
\includegraphics[width=0.7\textwidth]{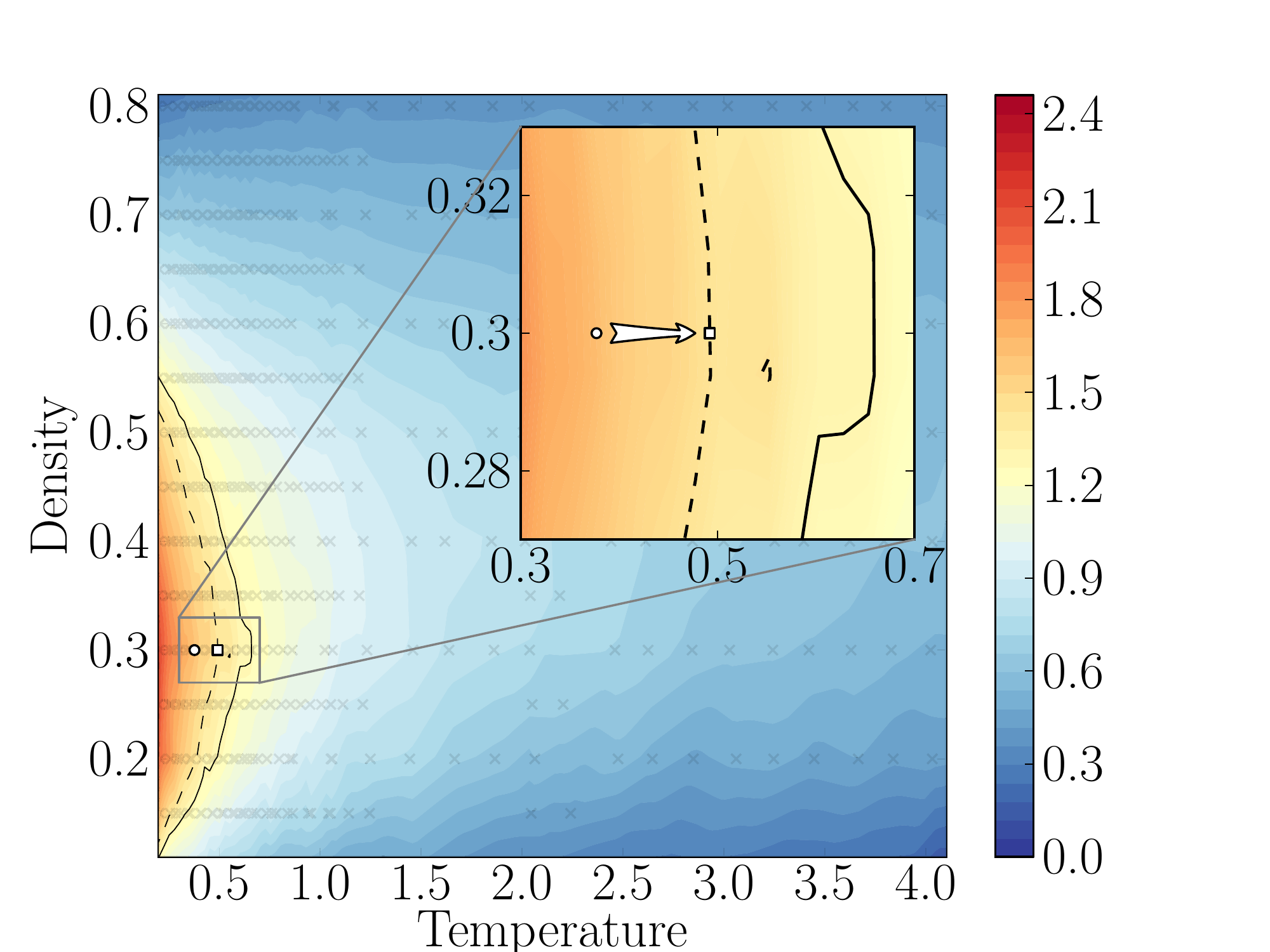}
\caption{Contour of $1/\nu_{_{\text{MD}}}$ for a range of density and 
temperatures in MD units, interpolated from a parameter study of $660$ 
Couette channels (sample points shown by crosses).}  
\label{Re_study}
\end{figure}

The angular brackets should formally denote an ensemble average over many systems \citep{Irving_Kirkwood}, consistent with expressing continuum concepts as the expectation of molecular quantities \citep{Pope}.
In practice, provided the average phenomena evolves at a slower timescale than the molecular interactions, the angular brackets can be used to denote an average over some time interval, $\tau_{_{MD}}$, given by,
\begin{align}
\langle \mathcal{A}(x,y,z,t) \rangle \define \frac{1}{\tau_{_{MD}}} \int_0^{\tau} \mathcal{A}(x,y,z,t^\prime) dt^\prime 
\end{align}
Despite averaging over the time interval, the left hand side is still a function of a longer timescale.
This is a distinct difference to the laminar steady state molecular simulation typically studied by NEMD.
This results in the presence of at least two distinct time scales in a turbulent molecular flows.
To obtain the Reynolds stress tensor, we start by considering the sum over the outer product of individual molecular velocities. 
By separating out the streaming velocity, we obtain the kinetic part of the molecular pressure tensor in \eq{Stress} and a convective term, $\rho \boldsymbol{u} \boldsymbol{u}$, which can be further split into mean flow and turbulent fluctuations, \ \ie{} 
\begin{align}
\overline{\sum \langle m_i\dot{\boldsymbol{r}}_i\dot{\boldsymbol{r}}_i} 
\rangle =  \overline{\sum \langle \boldsymbol{p}_i 
\boldsymbol{p}_i/m_i  \rangle} + \overline{\rho \boldsymbol{u} \boldsymbol{u}
\vphantom{\boldsymbol{u}^\prime}}
+\overline{\rho  \boldsymbol{u}^\prime \boldsymbol{u}^\prime
\vphantom{\boldsymbol{u}^\prime}}
\label{Triple_decomp}
\end{align}
the angular brackets and overbar denote averaging over $1600$ and $5.6 \times 10^6$ timesteps respectively. 
Note that for simplicity, decomposition of the scalar pressure, $P = \overline{P} + p^\prime$, has not been considered here.

In order to ensure turbulence in the molecular channel, it is desirable to maintain $Re_{_{\text{MD}}} \geq 400 $ throughout the MD simulation. 
Determining a molecular viscosity which allows computationally tractable simulations is key to the work presented here: the lower the viscosity, the smaller the required MD domain for $Re_{_{\text{MD}}} \geq 400$. 
To this end, $660$ independent Couette simulations (domain size $15.8\ell \times 15.8\ell \times 15.8\ell$) were run at different density and temperature values with the ratio of stress and strain used to obtain the viscosity \citep{Travis_Gubbins_00}. 
The summary of this study is shown as a contour of $1/\nu_{_{\text{MD}}}$ in Figure \ref{Re_study}. 
A density of $0.3$ with low system temperatures can be seen to give the lowest viscosity. 
As a result, a density of $\rho_{_{\text{\text{MD}}}}^{\text{fluid}}=0.3$ and temperature $T=0.4$ were chosen so $1/\nu_{_{\text{MD}}} \approx 1.6$.
This requires a channel half height of $h=280.4\ell$ (with $3\ell$ wall, $283.4\ell$ total) resulting in around $300$ million molecules ($N = 291,287,810$).
The MD domain size was therefore $1560.4 \ell \times 566.7 \ell \times 1069.9 \ell$ and the expected Reynolds number $Re = hU_w/\nu_{_{\text{MD}}} \approx  450$. 
The final choice of domain size is somewhat arbitrary, aiming to maximize the simulated Reynolds number, ensure wall velocity, density and temperatures are within the range of previous NEMD studies \citep{Ahmed_Sadus} and still present a computationally tractable simulation.

The wall velocity, $U_w = \pm 1$, and consequent strain rate, $\dot{\gamma} = 0.0035$, are small compared to molecular dynamics studies \citep{Evans_Morris}.
This is in order to minimize non-linearity in viscosity coefficients, shear heating and compressibility effects. 

The temperature $T=0.4$ is applied as a thermostat target value to the solid walls which have a density $\rho_{_{\text{MD}}}^{\text{solid}}=1.0$.
The initial temperature, $T=0.4$, is set by randomly choosing velocities for molecules in the face centered cubic lattice.
In the liquid region, the molecules are initialized at the same density as the walls and removed to get the density of $0.3$. 
On initialization, the initial structure melts and the average domain temperature at the start of the simulation drops to $T \approx 0.377$.
As in the continuum solver, a Reynolds number of around $400$ is sub-critical and a careful choice of initial condition to ensure turbulent flow is required \citep{jimenez_moin}.
Having setup the domain, the molecular velocities were adjusted to impose the starting velocity field.
The same initial velocity field as used in the CFD run, obtained from an initial random perturbation at higher Reynolds number and steadily reduced using Channelflow, is applied on a cell by cell basis (uniform $64 \times 256 \times 64$ cells with $\sim 286$ molecules per cell). 
An adaptation of the algorithm to adjust mean velocity on initialization was employed \citep{Rapaport}.

The MD simulation was run with a timestep $\Delta t = 0.005$ for $5.6 \times 10^6$ timesteps, corresponding to the $100 h/U$ units required for a single regeneration cycle \citep{Hamilton}. 
Further runs for another $3.7 \times 10^6$ timesteps were employed to ensure the observed behavior was repeated.
This simulation required approximately $280$ thousand CPU hours, run over either $256$ cores ($8$ core Nehalem CPUs), $360$ or $720$ cores (Westmere $12$ core processors).
Inter-core communication is facilitated by the message passing interface (MPI) \citep{MPI1_book}.
The Imperial College London computing cluster, \texttt{CX2} was used with CPU employed instead of GPUs.
Current GPU random access memories are typically too small to hold the $300$ million molecules required for the current system.
The latest version of GPU based MD codes, such as HOOMD \citep{Anderson_et_al}, are beginning to allow distribution over multiple GPUs (through MPI style communications).
Together with increasing GPU random access memories, this may soon allow large runs of the type presented here to be routine on desktop computers.

Due to the work done by the sliding walls in a Couette flow simulation, heat is added to the MD system, which increases the viscosity and decreases the Reynolds number. 
From the parameter study, the channel's Reynolds number is approximately proportional to the inverse square root of temperature, $Re(T) \approx 300 /\sqrt{T} $.
At the initial temperature of $T \approx 0.377$, the starting Reynolds number is $Re_{_{\text{MD}}}=489$.
The arrow in Figure \ref{Re_study} shows the change during a single flowthrough time ($5.6 \times 10^6$ timesteps) due to heating, with a final value of $Re_{_{\text{MD}}}=424$. 
The minimum Reynolds number for which turbulence is sustained, $Re=360$ is shown as the solid black line in Figure \ref{Re_study} and the CFD Reynolds number, $Re_{_{\text{CFD}}}=400$, is shown by a dotted line. 
The choice of parameters therefore ensured that $Re_{_{\text{MD}}}$ was greater than $400$ for the regeneration cycle.
Even during a second regeneration cycle, the temperature increased to $T = 0.52$ and the Reynold number is still $Re \approx 416$.

\subsection{Setup Summary}

The length and viscosity in the CFD simulation are matched to the molecular simulation for simplicity.
The key parameters of both CFD and MD models are presented in table \ref{table} for comparison
\begin{center}
\begin{tabular}{|c|c|c|}\hline
Quantity & CFD &  MD \\\hline
$\; Lx \times Ly \times Lz \;$ & $\; 1.75\pi h \times 2h \times 1.2 \pi h \;$ &  $1560.4 \ell \times 566.7 \ell \times 1069.9 \ell$ \\\hline
Simulation  & Gauss-Lobatto in $y$ & Nmols: $291,287,810$  \\ 
Size & $ 64 \times 256 \times 64 $ & Uniform Grid: $ 84 \times 198 \times 50 $ \\\hline
$T$ & Isothermal & Variable $0.377$ to $0.500$  \\\hline
$\Delta t$ & Variable, max $8$ & $0.005$ \\\hline
$\tau_{_{ave}}$ & $28,000$ & $8$ \\\hline
$Re$ & $400$ & Variable $489$ to $424$  \\\hline
\end{tabular}
\label{table}
\end{center}
Both CFD and MD Reynolds numbers are based on channel half height, $h=280.4\ell$.
The MD simulation has walls of thickness $3$ on both top and bottom, leaving a fluid region of height $560.7\ell$. 
The simulation size row contains the grid points for the CFD case and number of molecules (with typical averaging bins) for the MD case.
The averaging time, $\tau_{_{ave}}$ is used to define the averaging period in the overbar and angular brackets definition of mean flow in the CFD and MD respectively.
Both simulations write velocity output files at the same frequency (every $8$ time units).
This is a snapshot for the CFD while the MD field is obtained from an average of $64$ bins in time, each separated by $25$ timesteps to ensure decorrelation.

In the next section, the results of from these two models are compared.

\section{Results}
\label{Sec:results}

In this section, the results from the molecular simulation are compared to the same case simulated using computational fluid dynamics (CFD). 
In order to verify the behavior is turbulent; first, instantaneous structures are shown to be consistent with the CFD solution and the key features of the turbulent regeneration cycle are shown to be correctly captured by molecular dynamics.
Next, the dynamic behavior of the system is demonstrated by evaluating the input energy vs dissipation phase diagram.
The evolution of various forms of energy are presented and the impact of increasing temperature is discussed.
Time averaged statistics are presented and verified, both against the CFD results and studies from the literature.
The law of the wall is shown with high resolution velocity averaging near the wall to highlight the expected stick-slip behavior between the MD wall and fluid.

Next, molecular results are presented, with particular emphasis on the unique insights offered by this discrete nano-scale modeling technique.
The impact of the size of averaging volume is explored, by sub-dividing the domain into different averaging volumes.
The velocity energy spectra in the CFD, laminar and turbulent MD systems are compared, with a discussion of the possibility of sub-grid motions, viscous dissipation and the minimum scale of turbulence.
The probability density functions of velocity in a molecular system are evaluated showing the range of fluctuations and the relative subtlety of turbulent behavior.
Finally, a range of Lagrangian style statistics are presented as an example of the powerful insights offered by a molecular model.

\subsection{Instantaneous structures}
\begin{figure}
\includegraphics[width=\figwidth\textwidth]{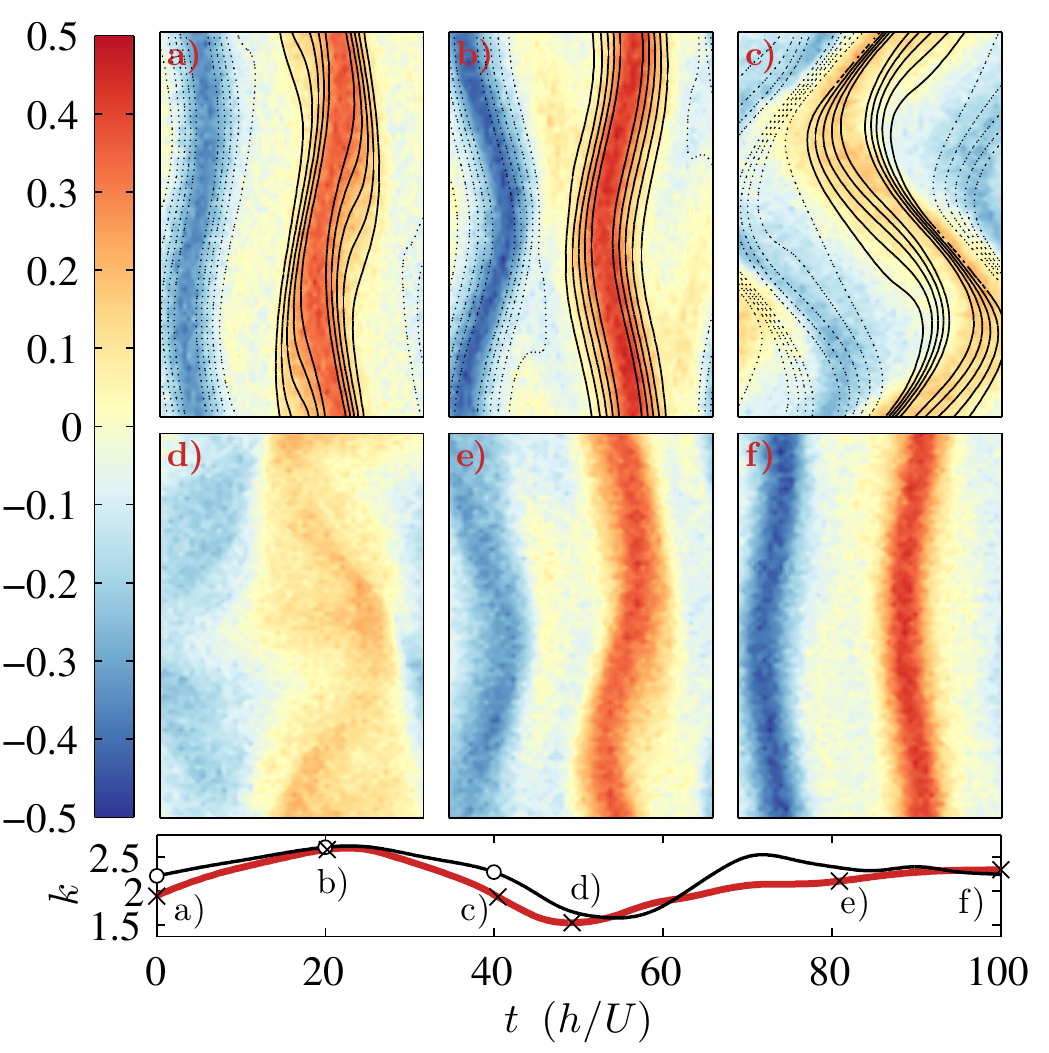}
\caption{Contour plots of $u$ velocity on the $xz$ plane at the centreline for MD 
(colors) and CFD (black contours with positive ($\xline{}$) and negative 
($\cdot \cdot \cdot$), separated by $0.1$). The times of the six contour plots, 
$a)$ to $f)$, are denoted on the plot (CFD ($\circ$) and MD ($\times$)) showing 
the evolution of whole domain turbulent kinetic energy $k$ (magnitude of $k$ 
is $10^3$ times display value).}
\label{Contours}
\end{figure}
In starting from the initial CFD solution, the expected bursting phenomena is reproduced over a whole cycle.
The turbulent streaks in the MD system can be clearly observed in Figure \ref{Contours}, as well as their breakdown and reformation. 
The MD velocity fields, $\boldsymbol{u}$, was obtained by averaging the velocity of molecules on a uniform grid of $84 \times 198 \times 50$ cell with $64$ samples in time each separated by $25$ timesteps to avoid correlations \citep{Rapaport}. 
Figure \ref{Contours} shows the contours of velocity at the channel centreline in the MD system. 
The breakdown and reformation of the streaks is consistent with well documented behavior in the CFD literature \citep{Hamilton,smith2005low}. 
The overlayed black line contours in Figure \ref{Contours} $a)$ to $c)$ are the results from Channelflow at equivalent times. 
Qualitatively similar turbulent structure are clearly present in both the CFD and MD simulation. 
The molecules model includes both temperature and density variations.
 
The compressibility in the MD models would not be expected to be negligible, with speed of sound predicted to be approximately $248m/s$ \footnote{From the ideal gas equations, the predicted speed of sound is about  $130m/s (T=48K)$, although to take into account the effect of dense fluid effects, the speed of sound is measured by an initial shock wave observed on start-up which moves at $248m/s$}.
The wall is sliding at a speed of $160m/s$ so the the Mach number is therefore about $0.65$

%

It is therefore surprising that there is such good agreement between an isothermal and incompressible continuum solution and the molecular model.
It appears that despite significant spatial and temporal variation in temperature, the coupling to viscosity is weak and does not appear to effect the hydrodynamics.
Although there is good agreement for the first part of the regeneration cycle, the solutions diverge as time progresses and Figure \ref{Contours} $d)$ to $f)$ are shown for the molecular model only. 
This divergence is mainly attributed to the non-linear nature of both the MD and CFD flows.
Exact structures observed in two continuum cases started from the same initial condition would also diverge over time.
This time evolution of the instantaneous turbulent kinetic energy $k = \frac{1}{2} [ u^{\prime2} + v^{\prime2} + w^{\prime2} ]$, is shown below the contours in Figure \ref{Contours}.
The reformation of the streaks from the point of minimum turbulent kinetic energy, Figure \ref{Contours} $d)$, is clearly seen in the successive MD only contours and the plot of turbulent kinetic energy below Figure \ref{Contours}.

\begin{figure}
        \centering
        \begin{subfigure}{0.45\textwidth}
                \includegraphics[width=\textwidth]{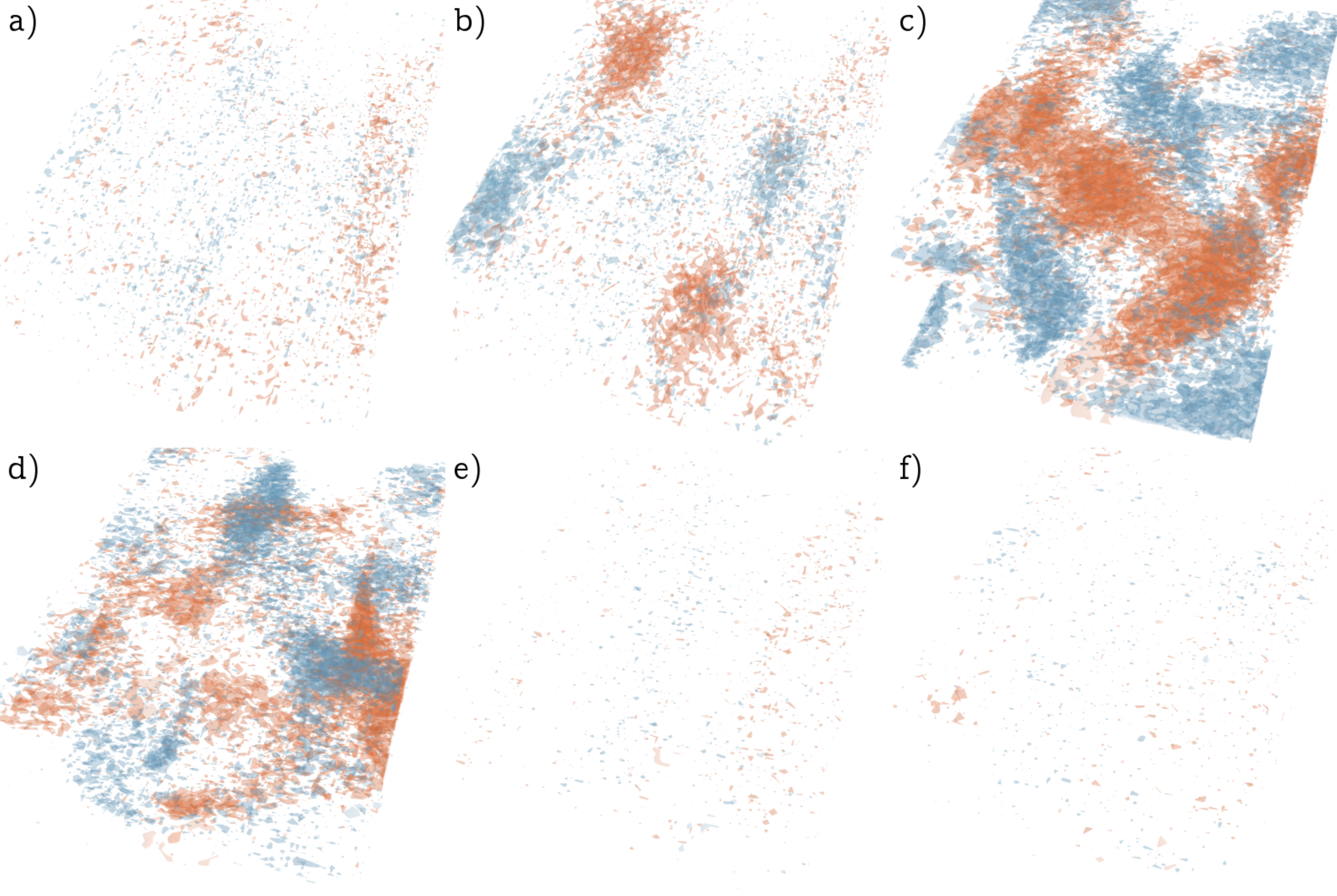}
                \caption{Isosurfaces of streamwise vorticity, $\omega_x=-0.005$ (red) and $\omega_x=0.005$ (blue).}
                \label{isocontours_vortx} 
        \end{subfigure} \;\;\; 
        \begin{subfigure}{0.45\textwidth} 
                \includegraphics[width=\textwidth]{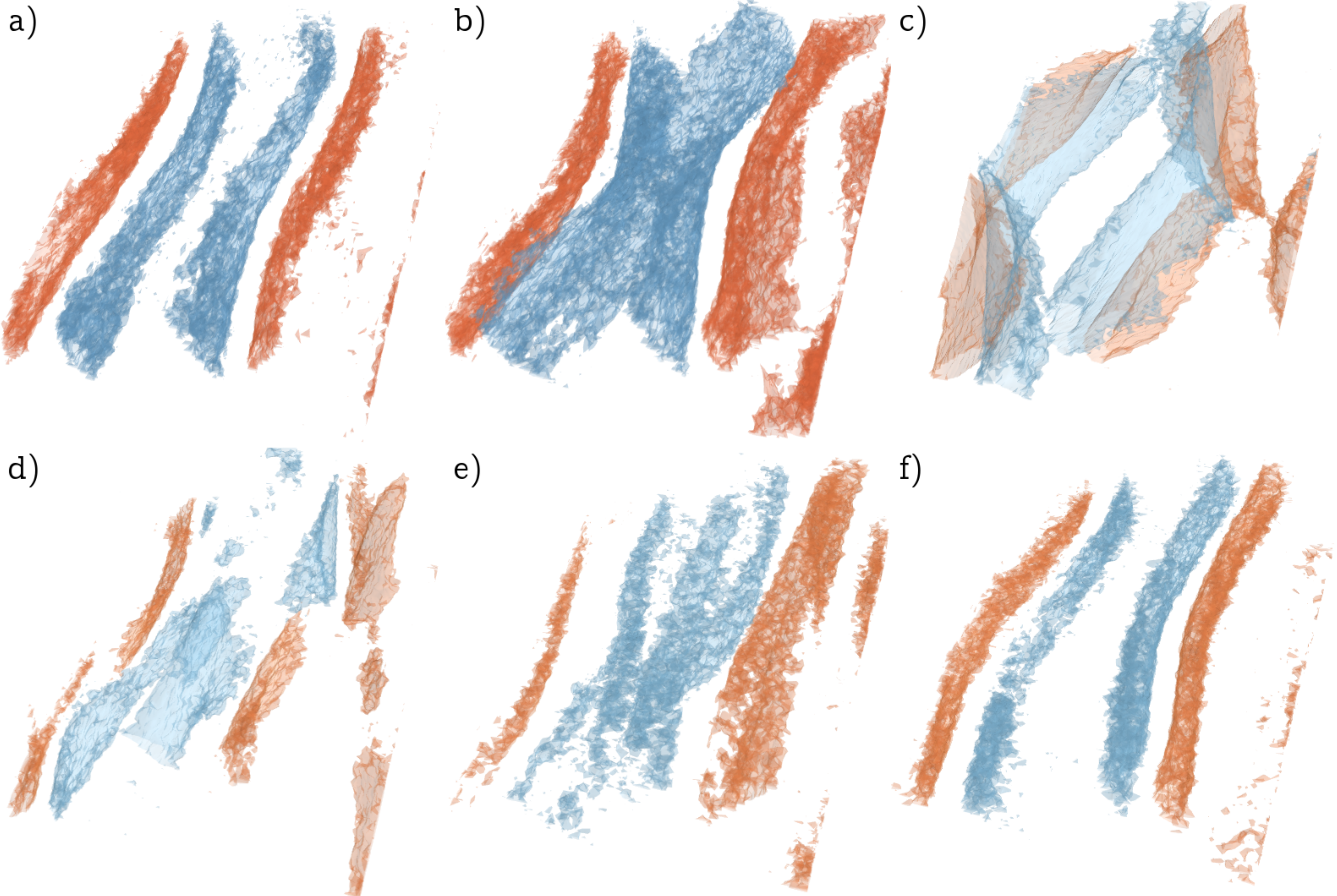}
                \caption{Isosurfaces of spanwise vorticity, $\omega_y=-0.0035$ (red) and $\omega_y=0.0035$ (blue).}
                \label{isocontours_vorty}
        \end{subfigure}
        \caption{Vorticity isosurfaces at times, $a)$ to $f)$, from figure \ref{Contours}.}
        \label{isocontours_vorticity}
\end{figure}
The cycle shown in Figure \ref{Contours} is an example the so called bursting phenomena, which includes the break down and reformation of coherent structures in turbulent flow \citep{Viswanath}.
This has been observed in both experimental \citep{Bech_et_al} and numerical studies of turbulence \citep{Hamilton,KAWAHARA_KIDA}. 
This cycle consists of three parts; first the streaks become unstable and break apart (streak break-down), next the stream-wise vortices in the flow become stronger (vortex regeneration), 
finally a new set of streaks form from the streamwise vortices (streak formation) \citep{Hamilton}.
The streak breakdown and reformation are clearly observed in Figure \ref{Contours}. 
%
%
%

\begin{figure}
\includegraphics[width=\figwidth\textwidth]{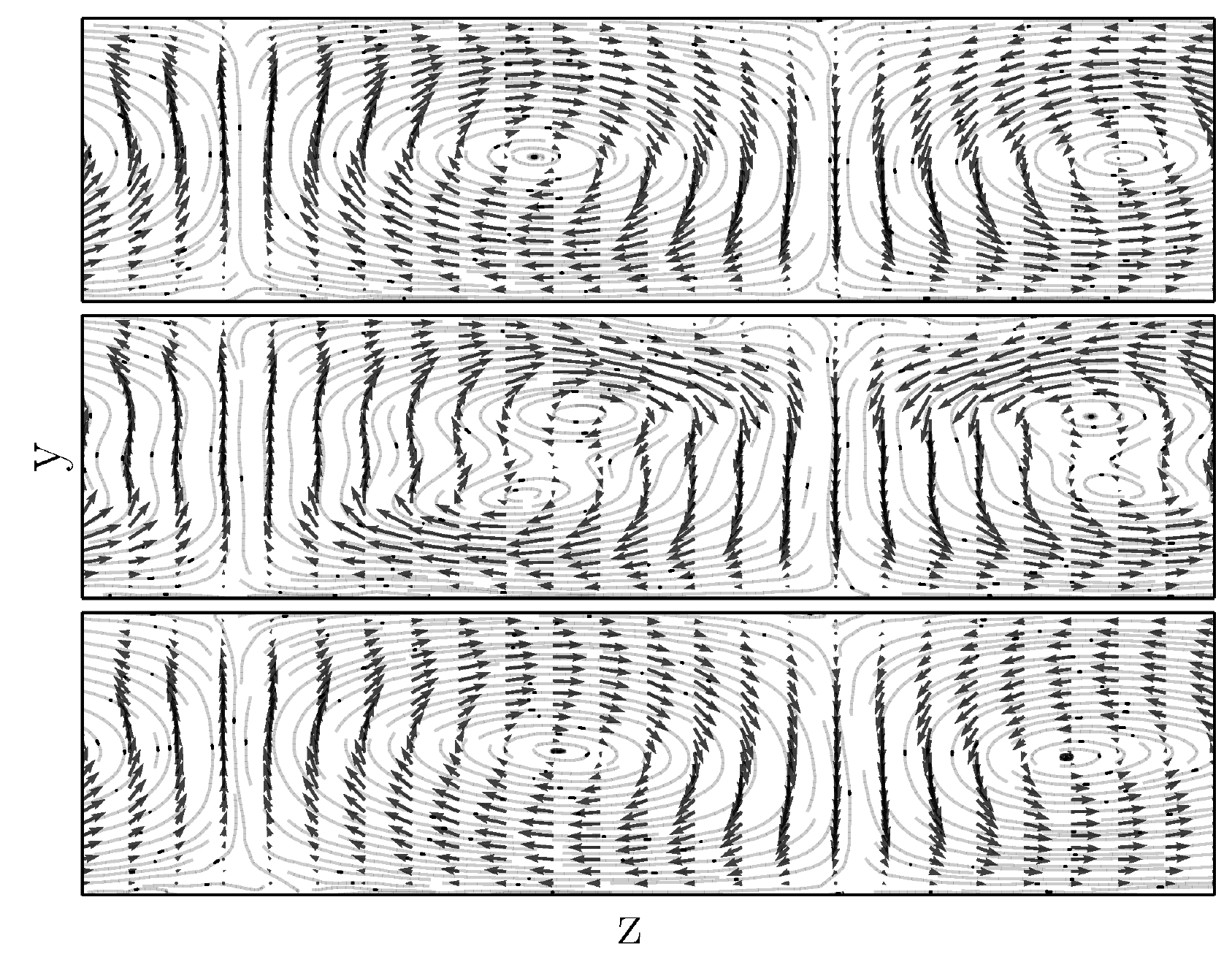}
\caption{$UW$ velocity vectors and streamlines on the $yz$ plane averaged over all $x$. Contour shows temperature variations relative to the mean of the whole domain
at that time. Times are chosen before streak breakdown, directly after breakdown and after streak reformation.}
\label{xz_Sections}
\end{figure}
The formation of vortices at the point of breakdown can be observed by looking at vectors of $v$ and $w$ in Figure \ref{xz_Sections}. 
By comparison with the vortices at the end of the simulation (when the streaks had reformed) it is clear that the vortices are much stronger at the point of maximum breakdown.
Indeed, a secondary vortex appears to have formed, which may be a result of the slightly higher Reynolds number at the time of breakdown ($Re \approx 430$) in the MD system.
The vorticity isosurface of Fig. \ref{isocontours_vorticity} show a large increase in streamwise vortices at the point of streak breakdown in the minimal channel cycle.

Figures \ref{Contours} and \ref{xz_Sections} suggest that the key stages of the near wall regeneration, or bursting, cycle are reproduced in a molecular simulation. 
This bursting phenomena is commonly believed to be the fundamental mechanisms of turbulent energy production \citep{Viswanath}. 
For larger systems, it has been suggested that turbulence cannot be maintained without these near wall bursting structures \citep{Hamilton, jimenez_moin}.
As this fundamental characteristic of turbulent flow can be reproduced by a purely molecular model, it could provide a new perspective in the simulation of turbulence flow.

As the initial condition used for the molecular system is based on a turbulence CFD field, it is possible that the observed behavior is a result of inertia or memory of that initial field. 
To further test the persistence of this MD regeneration mechanism, the molecular channel at $t=5.6 \times 10^6$ and $Re = 424$ is run beyond the single cycle.
A similar breakdown is observed at iteration $7.2 \times 10^6$, followed by vortex regeneration and subsequent reformation of the streaks.
It seems unlikely that such complex behavior could occur simply from an initial condition, especially over two distinct and unique regeneration cycles.

In experimental flows below a critical Reynolds number (about $360$) the breakdown of the streaks occur prematurely and they do not reform \citep{Bech_et_al}.
To explore this, two smaller MD simulations are run with the same initial condition, temperatures and density but smaller domains containing $2,048,308$ and $11,935,488$ molecules respectively.
The effective Reynolds numbers are $Re=40$ and $Re=100$ and in both cases, the streaks and vortices dissipate and a transition to laminar flow is observed.
This behavior is consistent with equivalent simulations performed using Channelflow at Reynolds numbers of $40$ and $100$.
The appropriate MD domain size for a Reynolds number greater than $400$ therefore observes at least two regeneration cycles, while in smaller domains, the streaks dissipate and the flow relaminarizes.

\begin{figure}
\includegraphics[width=0.7\textwidth]{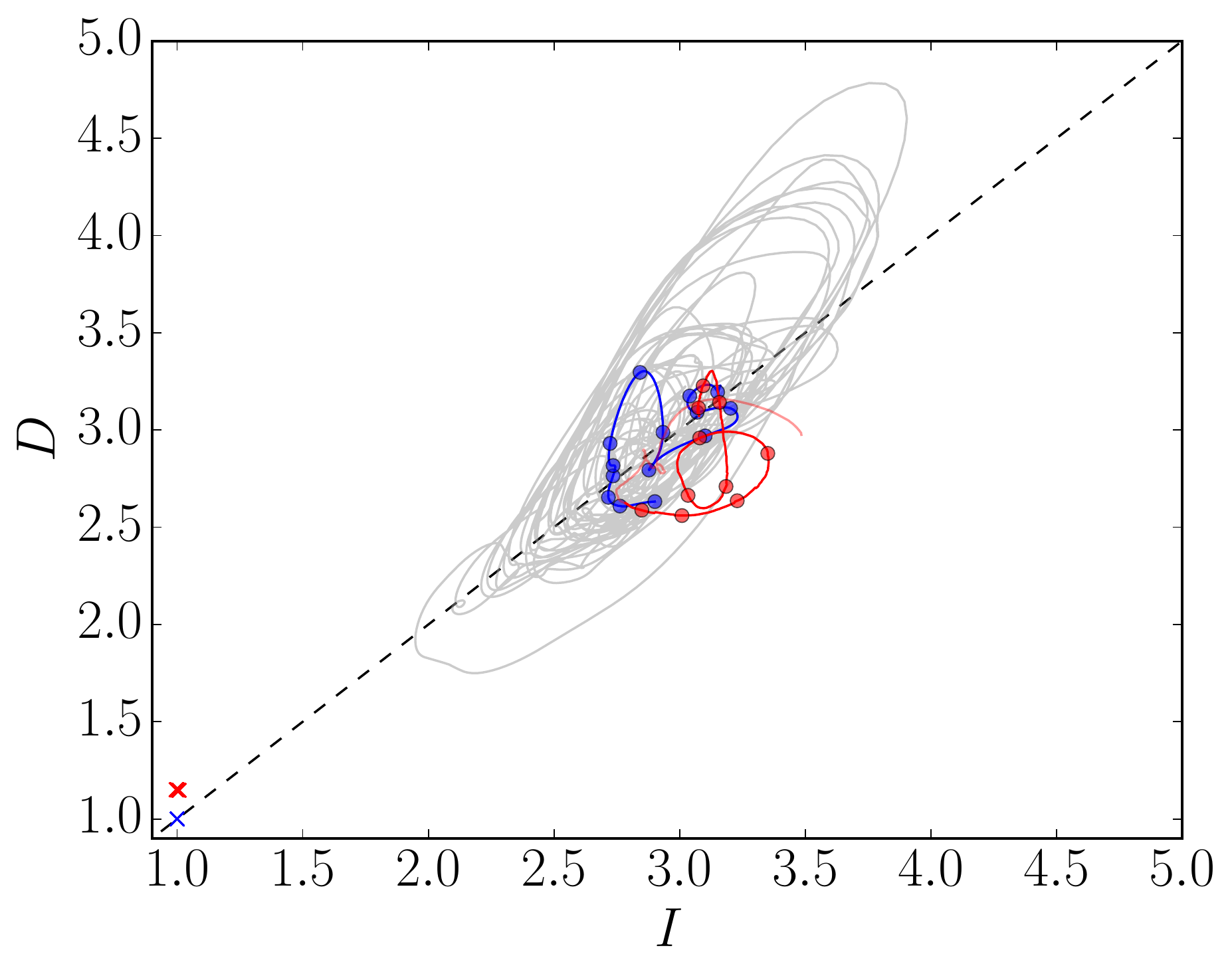}
\caption{Plot of energy input vs dissipation normalized by the laminar solution. The CFD trajectory over $35$ regeneration cycles (\textcolor{lgrey}{$-$}) and for the single flow through (\textcolor{blue}{$ - - \!\!\!\!\!\!\!\! \bullet \;$ }) compared to the MD single flow through (\textcolor{red}{$ - - \!\!\!\!\!\!\!\! \bullet \;$ }) and beyond (\textcolor{red}{--}). The laminar solutions in both cases are shown by (\textcolor{blue}{$\times$}/\textcolor{red}{$\times$}) for CFD and MD respectively}
\label{IvsD}
\end{figure}

The phase space plots of Figure \ref{IvsD} lend further weight to the assertion that the flow remains turbulent throughout the simulation. 
The dissipation is given by,
\begin{align}
D= \frac{1}{V} \int_V \left( |\boldsymbol{\nabla} u|^2 + |\boldsymbol{\nabla} v|^2 + |\boldsymbol{\nabla} w|^2 \right) dV
\end{align}
and the mechanical energy input at the wall is,
\begin{align}
I= \frac{1}{A} \int_A \left( \frac{\partial u }{ \partial y} \bigg|_{y=-h} + \frac{\partial u }{ \partial y} \bigg|_{y=h} \right) dA.
\label{I}
\end{align}
The MD values are averaged over $6400$ samples separated by $25$ time steps each in order to minimize the molecular scale fluctuations in the sample data.
In a steady state laminar flow, the $I$ and $D$ values are equal, denoted by the dotted line in Figure \ref{IvsD}.
Both CFD and MD are normalized by the analytical solution for the steady laminar Couette flow solution.
The numerical results for simulated laminar flow are shown at the bottom left of Figure \ref{IvsD}. 
The MD laminar solution is larger than unity due to the inherent thermal fluctuations (which has a greater impact on derivatives such as strain rates).

These phase plots are inspired by chaos theory style representations of dynamical system.
The many degrees of system are reduced to two central features for wall driven shear, namely the rate mechanical energy is input into the flow and the rate at which the mechanical energy is dissipated from coherent velocity to heat.
For the minimal flow unit, the various stages in the regeneration cycle see greater dissipation at some times and more energy input at others.
As the dissipation should approximately balance the input on average for a cycle, an orbiting behavior is observed in $I$ vs $D$ phase space.
The non-linear hydrodynamic behavior is similar in both the CFD and MD model, evidenced by the similar orbiting behavior in Figure \ref{IvsD} for both systems.

Note, however, that the MD dissipation is observed to be lower on average than the CFD.
The CFD system is in a thermodynamic steady state with a constant viscosity while as the molecular system progresses the temperature and viscosity continues to change.
The dissipation $D$ and input $I$ only measure the energy change in hydrodynamic velocity components.
Energy added to hydrodynamic components by the wall, $I$, will predominantly change to thermal energy as measured by the mechanism for dissipation $D$. The two therefore appear to still balance in the MD system. 
The sliding and thermostatting in the molecular wall will exchange energy directly with the thermal components of the MD system, but these do not appear to significantly impact the $I$ vs $D$ cycle.

To further explore the energy change in a molecular system, the evolution of various forms of energy is considered next. 
The energy in the system can be divided as follows,
\begin{subequations}
\begin{eqnarray}
\mathcal{K}_{flow} & = & \frac{1}{2} \int_V  \rho |\boldsymbol{u}|^2 dV, \label{Kflow} \\
\mathcal{K}_{Them} & = & \frac{1}{2} \sum_{i=1}^N m_i |\boldsymbol{p}_i|^2 =  \frac{1}{2} \sum_{i=1}^N m_i |\boldsymbol{v}_i - \boldsymbol{u}|^2, \label{Ktherm} \\
\mathcal{K} & = & \frac{1}{2} \sum_{i=1}^N m_i |\boldsymbol{v}_i|^2  =  \mathcal{K}_{Them} + \mathcal{K}_{flow}, \label{K} \\
\mathcal{T} & = & \frac{1}{2} \sum_{i=1}^N \sum_{j=1}^N \phi_{ij} \label{T}, \\
\mathcal{H} & = & \mathcal{K} + \mathcal{T} \label{H}.
\end{eqnarray}
\end{subequations}
These include the total system kinetic energy, $\mathcal{K}$, separated into the the flow mechanical energy $\mathcal{K}_{flow}$ of the streaming velocity $\boldsymbol{u}$ (averaged over $64$ uncorrelated samples) and thermal energy $ \mathcal{K}_{Them}$. 
The potential energy, $\mathcal{T}$, is obtained from the sum of intermolecular potential interactions.
The total energy $\mathcal{H}$ (Hamiltonian) is the sum of kinetic and configurational contributions. 


\begin{figure}
\includegraphics[width=0.9\textwidth]{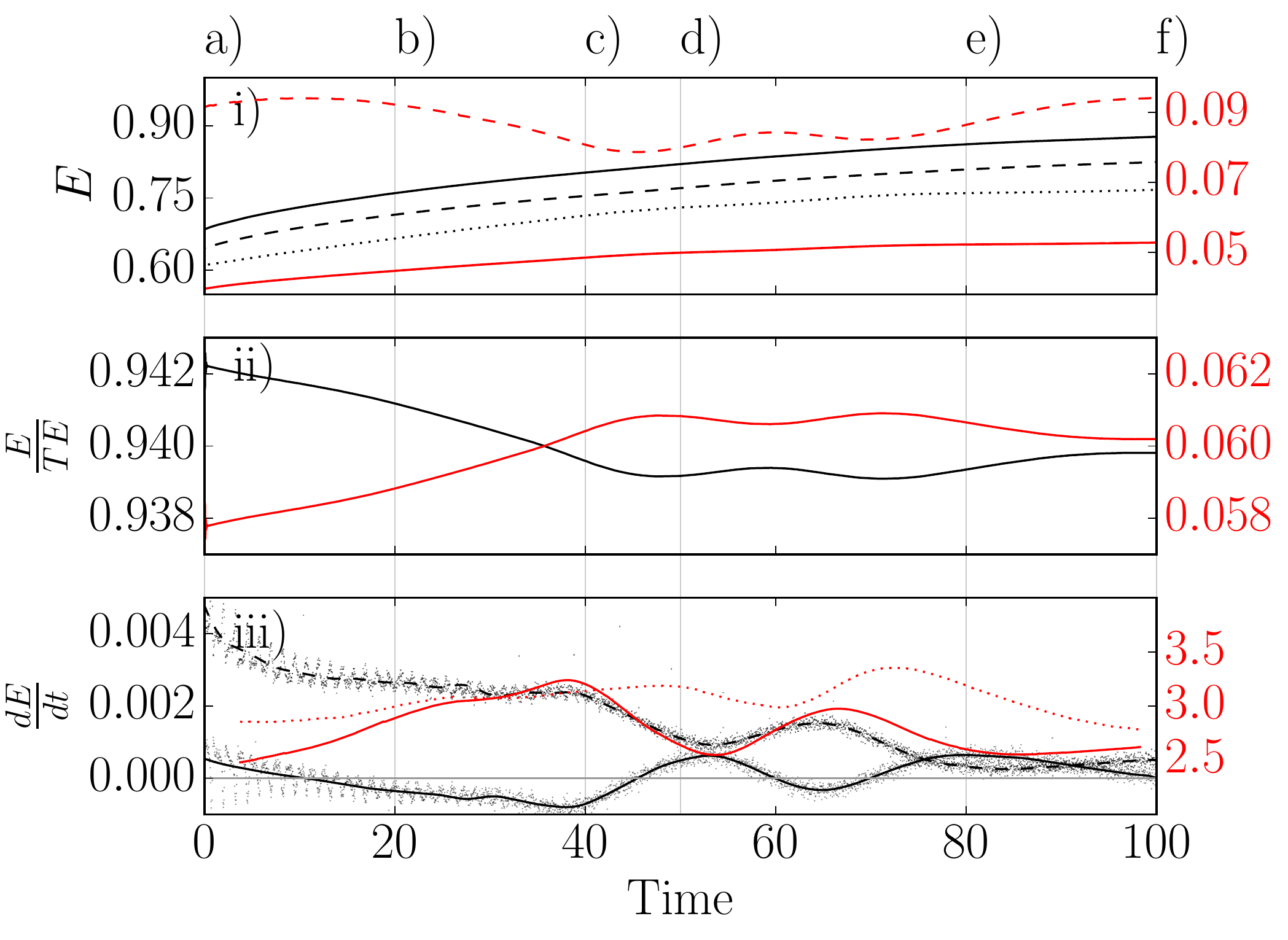}
\caption{ Plots of the energy changes as a function of time ($h/U$) in the molecular system, labelled with the cases from Fig \ref{Contours}. \\ i) Total energy, $\mathcal{H}$ ($-$), total system kinetic energy, $\mathcal{K}$ (- \!\!\! -), thermal energy, $ \mathcal{K}_{Them}$ ($\cdots$) with potential energy, $\mathcal{T}$ (\textcolor{red}{$-$}) and flow mechanical energy, $ \mathcal{K}_{flow}$ (\textcolor{red}{- \!\!\! -}) on second axis. \\ ii) Total kinetic energy, $\mathcal{K}/\mathcal{H}$, and potential energy, $\mathcal{T}/\mathcal{H}$, normalized by total energy shown on different axis with same range. \\ iii) Rate of change of flow mechanical energy, $\dot{\mathcal{K}}_{flow}$ ($-$) and thermal kinetic energy, $\dot{\mathcal{K}}_{Therm}$ (- \!\!\! -) from spline derivatives with actual points shown (\textcolor{llgrey}{$\boldsymbol{\cdot}$}), dissipation, $D/D_{laminar}$ (\textcolor{red}{$-$}) and energy input by shear, $I/I_{laminar}$ (\textcolor{red}{$\cdots$}) on the second axis.}
\label{energyhist}
\end{figure}

Figure \ref{energyhist}i shows the increase in the various types of energy, Eqs.\ (\ref{Kflow}--\ref{H}), in the channel.
The total system energy, $\mathcal{H}$ monotonically increases as the energy added by shearing is greater than the heat removed by the thermostatting in the walls.
The total kinetic and thermal energy monotonically increases, while the mechanical energy in the flow changes based on the regeneration cycle and appear to stay fairly constant.
There is an apparent decoupling between the thermodynamics and hydrodynamics energies.
Most CFD simulations do not observe significant heating \citep{Pope}.
In an MD simulation, the impact of heating is far more pronounced due to smaller system sizes and much higher shear rates.
The explicit energy conservation in the fluid and the choice to apply a thermostat to only the wall molecules results in the continual heating observed.
The heating has an impact of the effective viscosity in the molecular system which increases as the simulation progresses. 
Despite being away from thermodynamic equilibrium, the impact of this changing viscosity is not sufficient to prevent the regeneration cycle.


Figure \ref{energyhist}ii shows the evolution of normalised system kinetic energy, $\mathcal{K}/\mathcal{H}$, and potential energy, $\mathcal{T}/\mathcal{H}$.
By normalising using the total energy $\mathcal{H}$, the impact of the increasing system energy is removed and it is clear that a decrease in kinetic energy of the molecules results in a corresponding increase in potential energy.
The interchange of potential and kinetic energy is apparently effected by the breakdown and regeneration cycle with corresponding peaks and troughs (\textit{c.f.} \ref{Kflow} in Fig \ref{energyhist}i).
Higher potential energy is indicative of clustering of molecules. 
The hydrodynamic bursting cycle governs the flow of molecules and will result in changes in the density (for example inside the streaks).

Figure \ref{energyhist}iii shows the time evolution of flow energy, $\dot{\mathcal{K}}_{flow}$, and thermal energy $\dot{\mathcal{K}}_{Them}$.
The lines in the bottom panel Figure \ref{energyhist} are obtained from the derivative of a spline fit to the data (actual points in gray).
In addition, the input, $I$ and dissipation $D$ as plotted in the phase plot of Figure \ref{IvsD} are included.
The dissipation, $D$, closely matches the change in $\dot{\mathcal{K}}_{Therm}$ (after the initial part of the simulation), suggesting the dissipated mechanical energy is transferred directly to the thermal energy.
The time evolution of temperature is always positive as heat is continuously added to the system, however the rate is decreasing throughout the simulation.
In the extended run, not shown, by the end of the second regeneration cycle (approximately $160h/U$) the change in thermal energy appears to be zero on average.

There is a large decrease in the rate of thermal energy production from around the times c) to d), which coincides with the streak breakdown and vortex regeneration. It has been suggested that after streak breakdown, the vortices are re-energised in a complicated mechanism, requiring the interaction of several Fourier modes \citep{Hamilton, Waleffe}. The regeneration of the mean flow takes energy from smaller wavelengths and it is possible that some of the re-energization of vortices may be at the expense of what would be considered to be sub-grid thermal motion. The division between thermal energy and hydrodynamic energy is arbitrary in MD; dependent on the spatial and temporal averaging scales. The regeneration of the vortices could therefore be at the expense of energy which contributes to either thermal or hydrodynamic energy. A full energy analysis of the MD flow would be required to provide insight into distribution of energy in various forms.

\subsection{Time Averaged Statistics}
\begin{figure}
\includegraphics[width=\figwidth\textwidth]{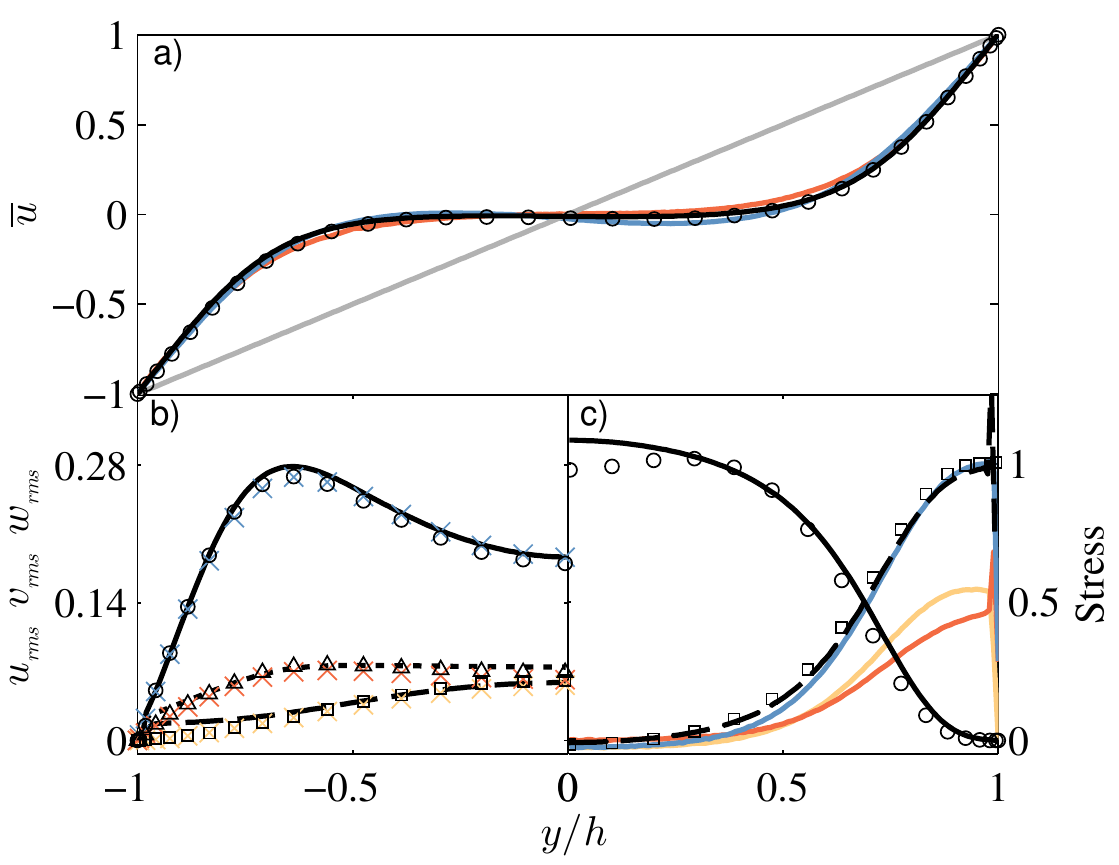}
\caption{Comparison plots of channel statistics for MD and CFD: a) mean  ($\xlinethick{}$), first (\textcolor{cm_red}{$\xlinethick{}$})
and last (\textcolor{cm_blue}{$\xlinethick{}$}) velocity profile for MD, mean CFD ($\circ$) and laminar analytical 
(\textcolor{lgrey}{$\xlinethick{}$}).
b) root mean square turbulent velocities 
$u_{_{\text{RMS}}},v_{_{\text{RMS}}},w_{_{\text{RMS}}}$ with MD 
($\xlinethick{},\xdashthick{},\boldsymbol{\cdot \cdot \cdot}$), CFD literature 
\citep{KAWAHARA_KIDA} ($\square,\circ,\triangle$) and Channelflow 
(\textcolor{cm_orange}{$\times$},\textcolor{cm_red}{$\times$},
\textcolor{cm_blue}{$\times$}) c) shear and turbulent stresses with MD/CFD 
for $(1/Re) d\overline{u}/dy$ ($\xdashthick{}$/$\square$), 
$\overline{u^\prime v^\prime}$ ($\xlinethick{}$/$\circ$), MD kinetic pressure 
(\textcolor{cm_red}{$\xlinethick{}$}), configurational stress 
(\textcolor{cm_orange}{$\xlinethick{}$}) and total pressure 
(\textcolor{cm_blue}{$\xlinethick{}$}). MD and CFD stresses in plot c) are normalized 
by MD and CFD viscous wall stresses respectively.}
\label{turbstats}
\end{figure}

In this section, we compare the time averaged statistical properties, both to published data and results from CFD simulations using Channelflow. 
Figure \ref{turbstats} compares a range of time averaged statistics as a function of wall normal position.
Panel \ref{turbstats}a) shows average velocity profiles for both the CFD and MD cases.
There is good agreement between the streamwise velocity $\overline{u}$, averaged over  $100h/U$ for the MD system compared to a very long simulation for the CFD ($3500h/U$). 
The straight line is the analytical solution for fully-developed laminar Couette flow. 
The flow is driven by the shearing due to the wall, which in a continuum model results in a repeating cycle at this Reynolds number.
The CFD model was run for over $5500$ flow through times with no sign of re-laminarization in the continuum case. 
As molecular dynamics is inherently energy conserving, shearing from the molecular walls results in a continual increase in the temperature of the molecular fluid.
Some heat energy is removed at the wall by a thermostat, but the available surface to remove heat is insufficient for a domain of this size.
The initial and final profile are displayed in Figure \ref{turbstats}a) for the molecular channel. 
Despite the decreasing Reynolds number, the profile still suggests the flow is turbulent.
The shape of this profile is due to the streamwise vortices in the flow \citep{Hamilton}, which as observed in Figure \ref{xz_Sections}, are clearly present throughout.
For the smaller molecular channels tested at effective $Re=40$ and $Re=100$, the turbulent streaks decay and the flow quickly relaminarizes, returning to the linear profile.

Root mean square (RMS) turbulence intensities for each velocity component are shown in Figure \ref{turbstats}$b)$ for both the CFD and MD systems. 
There is good agreement for the three velocity components, despite the stick-slip behavior at the walls, compressibility effects, temperature variations and the varying Reynolds number in the MD simulation.
In addition, results from the CFD literature\citep{KAWAHARA_KIDA} at $Re=400$ are compared in Figure \ref{turbstats}$b)$ demonstrating good agreement.
 
The shear and turbulent stresses are presented in Figure \ref{turbstats} c). 
Again, good agreement is observed between continuum and molecular results. 
The greater value of turbulent shear stress towards the channel center in Figure \ref{turbstats} c) is attributed to the higher Reynolds number of the molecular simulation.
Molecular dynamics does not require a viscosity to be defined, with pressure obtained directly from \eq{Stress} as a function of the molecular configuration and kinetic motions.
The molecular kinetic pressure and configurational stress contributions are shown in Figure \ref{turbstats} c). 
These MD pressure results are obtained on the same grid as the velocity measurements by recording every single interactions and crossings in a $1600$ timesteps period. 
During the simulation, $1500$ of these periods ($2.4 \times 10^6$ measurements in total) are taken at representative intervals.
The average kinetic pressure and configurational stress sum to a total shear pressure.
This total shear pressure can also be calculated in both the MD and CFD cases using $(1/Re) d\overline{u}/dy$. 
For the CFD case, $Re_{_{\text{CFD}}} = 400$, while for the MD case a Reynolds number of $Re_{_{\text{MD}}} = 430$ gives the best agreement with the molecular pressure.
An MD viscosity can be obtained from the average over the changing viscosity, $\mu_{_{MD}}$ using temperature and density to lookup values from the data in Figure \ref{Re_study}.
The viscosity which gives the best fit is approximately 10\% lower than predicted from Figure \ref{Re_study}. 
This may indicate that the turbulent flow in the molecular simulation has an impact on the viscosity coefficient.

Another interesting observation is the relative magnitude of the kinetic and configurational parts of the shear pressure.
Despite the low density of the fluid, the configurational shear is a comparable magnitude to the kinetic shear pressure while the direct configurational pressure contributes only $\tilde 5\%$ of total pressure.

%

Another characteristic of turbulent flow is the law of the wall, obtained by expressing the velocity and wall normal position in dimensionless units $y^+=y/\delta_\tau$ and $u^+ = u / u_\tau$ where $u_\tau \define \tau_w / \rho$ and $\delta_\tau = \nu / u_\tau$.
In dimensionless wall units, the observed behavior is similar for a range of different Reynolds numbers and flow geometry.
The definition of wall stress, $\tau_w$ is ambiguous in a molecular system as a result of stick slip behavior in the near wall fluid as well as deformation and stress in the wall itself.
The wall stress, $\tau_w$, is evaluated at a distance of $ 7 \ell$ away from the wall, the location at which shear stress equals the continuum analytical value in the laminar MD channel of the same dimensions. %
The stress is obtained using the method of planes definition, \eq{Stress}.

\begin{figure}
        \centering
        \begin{subfigure}{0.48\textwidth}
                \includegraphics[width=0.9\textwidth]{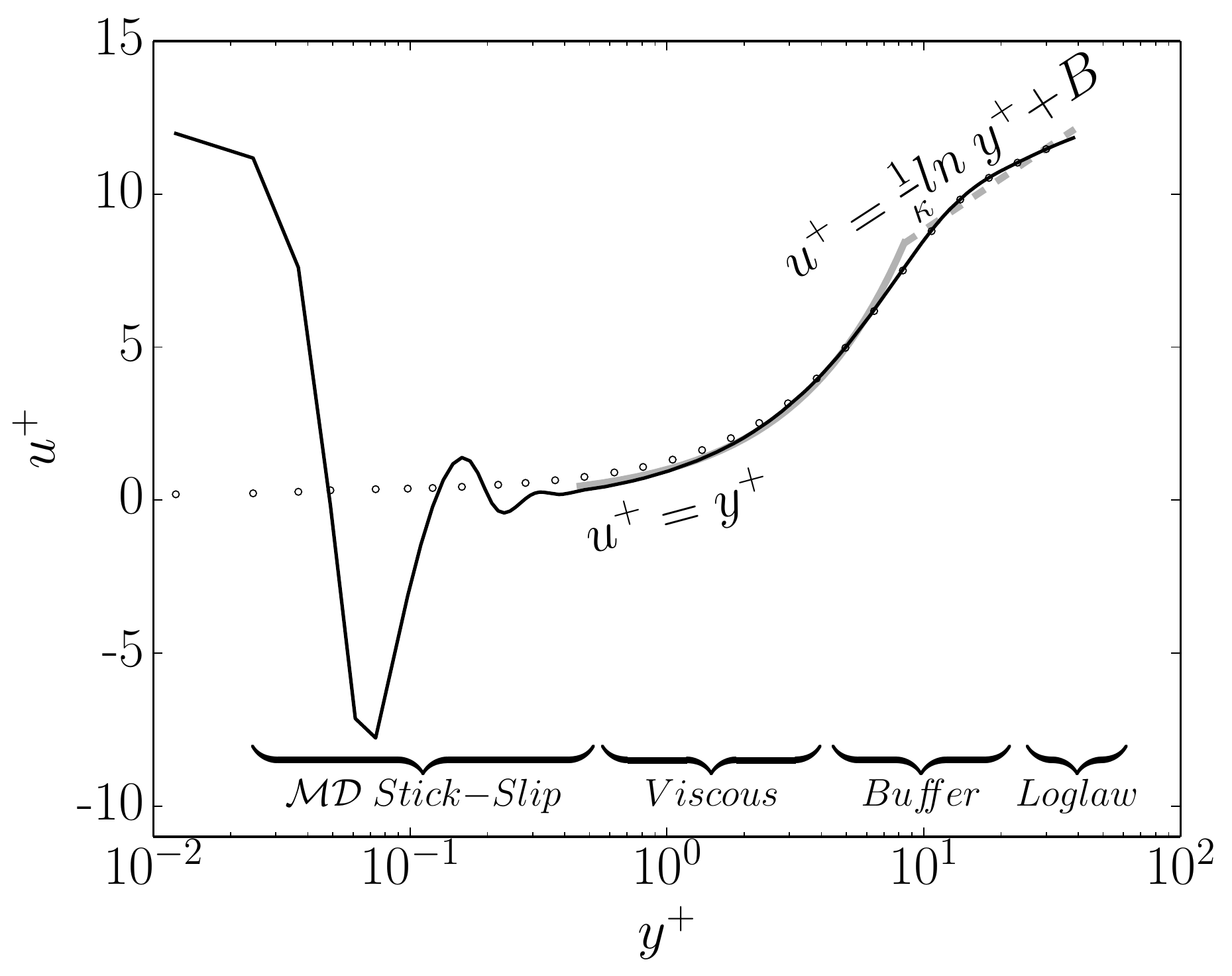}
                \caption{Law of the wall in reduced units over $3168$ $y$ cells shown for sum of cell momentum $\rho u / \rho_{mean}$ (\textcolor{black}{$-$}), cell velocity $u$ (logarithmically spaced $\circ$), viscous sub-layer analytical solutions (\textcolor{lgrey}{$-$}) and log law region (\textcolor{lgrey}{- -}).}
                \label{LOTW} 
        \end{subfigure}
        \begin{subfigure}{0.48\textwidth}
                \includegraphics[width=\textwidth]{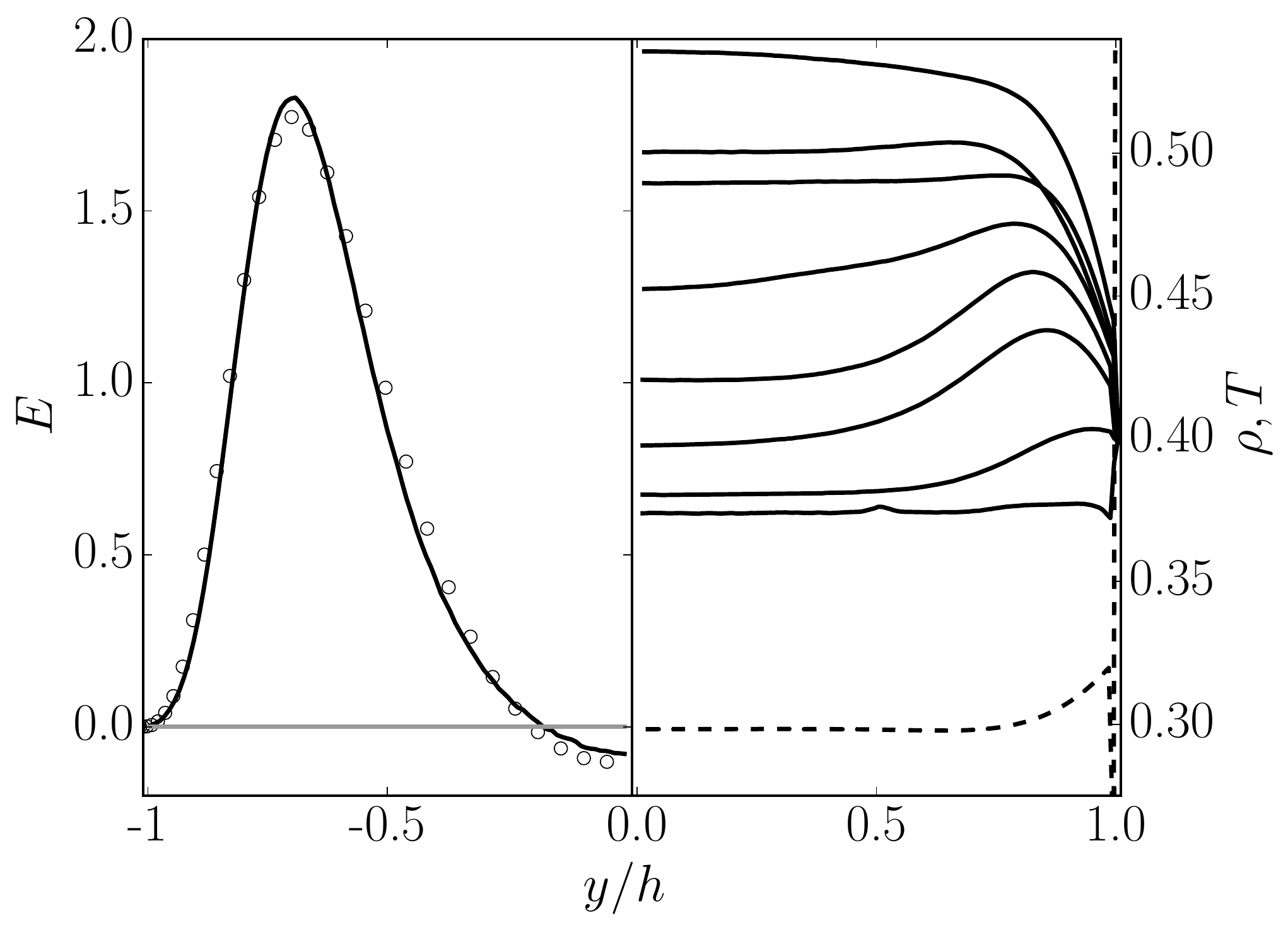}
                \caption{Turbulent production, $\mathcal{P}$, in the \\ MD (\textcolor{black}{$-$}) and CFD ($\circ$) case. \\ Right: Average density, $\rho$ (\textcolor{black}{- -}) \\  and temperature, $T$ (\textcolor{black}{$-$}) at times $t = \{0, 3, 14, 26, 43, 71, 100, 160\} h/U$. \\}
                \label{energychannel}
        \end{subfigure}
        \caption{Law of the wall and temperature profile in channel}
        \label{lawofthewall_and_energy}
\end{figure}

Figure \ref{LOTW} shows the velocity, $u^+$, against logarithmic wall normal position, $y^+$, both expressed in inner units.
The expected viscous sub-layer, and buffer regions are labeled.
The start of the log law region is also indicated but is not expected to be significant at such low Reynolds numbers \citep{Viswanath}.
Of main interest here is the molecular stacking region near the wall which is uniquely captured by molecular simulation. 
A high averaging resolution is employed, with $3168$ cell in the $y$ direction to explore the near wall detail.
This near-wall molecular stacking in Figure \ref{LOTW} has been widely observed in MD studies \citep{Travis_Gubbins_00} as well as experiments \citep{OShea_et_al}.
The fluid solid inter-molecular interactions result in stick-slip behavior and an effective Navier slip length.
It has been observed that wall roughness in a turbulent flow impacts the form of the viscous sub-layer and log law region \citep{Townsend}. 
Given the relative size of the wall features to the characteristic scales of the flow, the molecular roughness may impact the flow in nano and micro-scale channels. 
The law of the wall fit with surface roughness uses coefficients $\kappa = 0.41$ and $B=3.2$, perhaps suggests the impact of the molecular scale on the mean flow.
Higher Reynolds number MD simulations would be required to fully explore this.

The minimal channel models the evolution of turbulent structures in the near wall region.
This is where the majority of the production of turbulent energy occurs \citep{Kline_et_al}.
In Figure \ref{energychannel}, the rate of turbulent production, $\mathcal{P} \define \overline{u_\alpha^\prime u_\beta^\prime} d \overline{u_\alpha} / d r_\beta$ is compared for the CFD and MD simulations. 
There is good agreement between both modeling methodologies with similar profiles and maximum production at the same location.
The production appears to become negative near the channel centre in both cases, perhaps due to the low Reynolds number or insufficient statistics from a single flow through time.
On the left of Figure \ref{energychannel}, the density and temperature in the MD channel are also presented.
The density is seen to increase near the wall, which may be due to molecular stacking near the higher density solid ($\rho_{_{\text{MD}}}^{\text{solid}}=1.0$).
This may also be a consequence of the vortices in the flow, which move molecules near to the wall where they are slowed by the interaction with the wall.
The temperature is seen to be increasing throughout the simulation, while the wall temperature remains at $T=0.4$ due to the thermostat.
There is a Kapitza like jump near the wall which becomes more pronounced as the temperature difference between the fluid and solid becomes greater.
The initial temperature field is flat as only the velocity field is specified from the CFD initial condition.
The small peak in the initial temperature plot is a shock wave which appears due to initialization of a system between fixed walls.
This shock wave bounces between the walls and has dissipated by the next displayed time $3h/U$.
The effect of shear heating results in a continually increasing temperature and as a result, an increasing viscosity and decreasing Reynolds number.

In this section, the MD model has been compared to CFD results, showing good agreement for both the key stages of the regeneration cycle and reproduction of turbulent statistics.
Having demonstrated the molecular model is reproducing the minimal flow unit, in the next section the unique insights presented by turbulence in a molecular model are explored.

\subsection{Spectra}
\label{sec:spectra}
\begin{figure}
\includegraphics[width=0.98\textwidth]{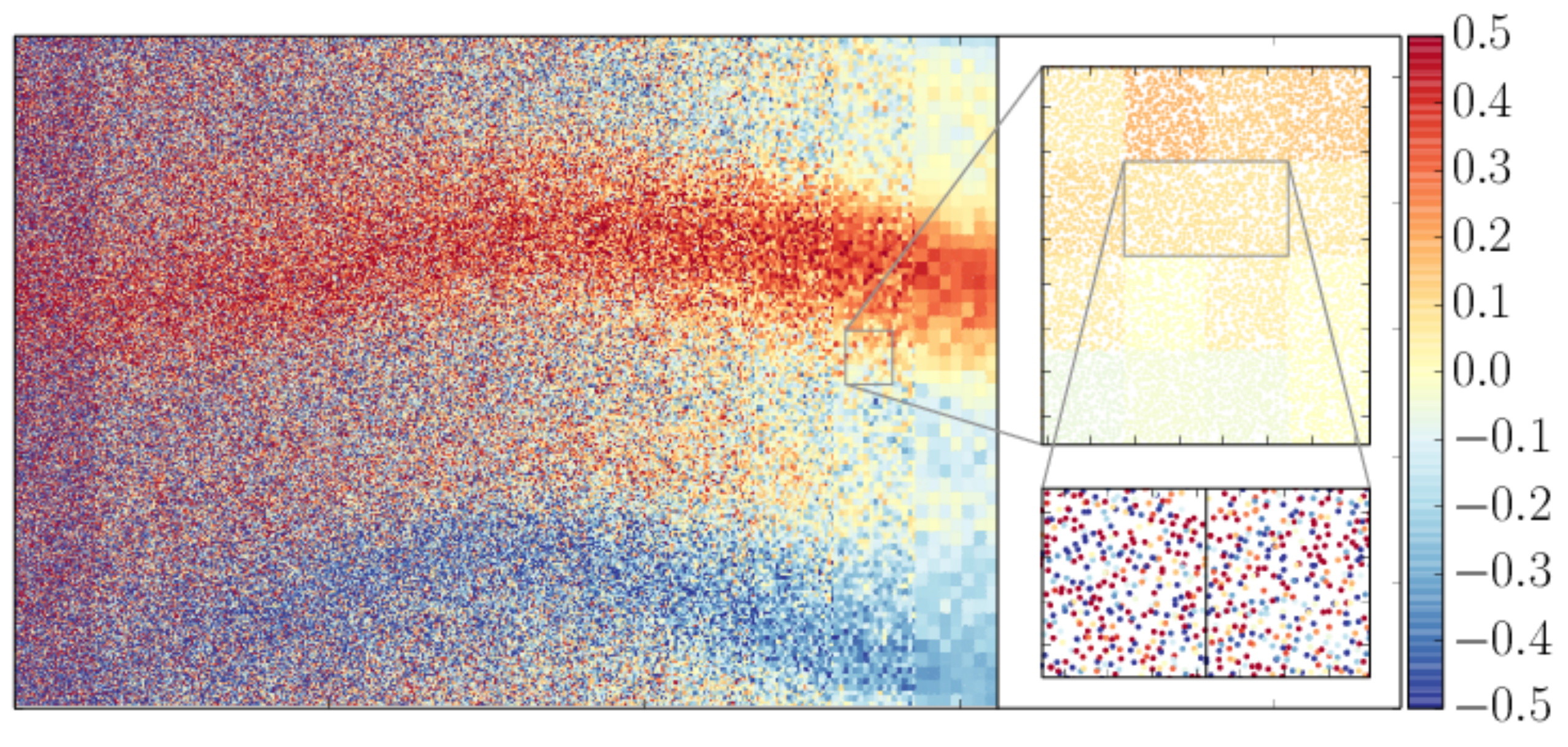}
\caption{Plot of $\sim 1.4$ million molecules in a central $y$ cells of width $2.86 \ell$. Coloring is by individual molecular velocity on the far left and moving right shows molecules colored by the average velocity in increasingly larger cells. The first insert is colored by cell velocities and the next level shows the velocities of individual molecules.}
\label{md_snapshotandzoom}
\end{figure}

A grid resolution study is performed in this section, with varying spatial and temporal averaging used to explore the flow structures at different scales. 
Figure \ref{md_snapshotandzoom} shows the impact of averaging on the observed velocity field at the center of the channel. 
The far left hand side shows individual molecules, each coloured by their own velocity.
Moving right, molecules are colored by the average velocity for the cell they are located in. 
The finest grained example has $800$ by $1344$ cells, with fewer cells employed in even steps until the far right velocity field is shown for the $84$ by $50$ cell resolution. 
It is apparent that only by employing appropriate spatial averaging can the details of the flow structures be observed.
To understand the impact of grid resolution on the energy content of the fluid flow, a Fourier transform is taken to obtain a spectrum of the various scales of fluid motion. 

\begin{figure}
\includegraphics[width=0.9\textwidth]{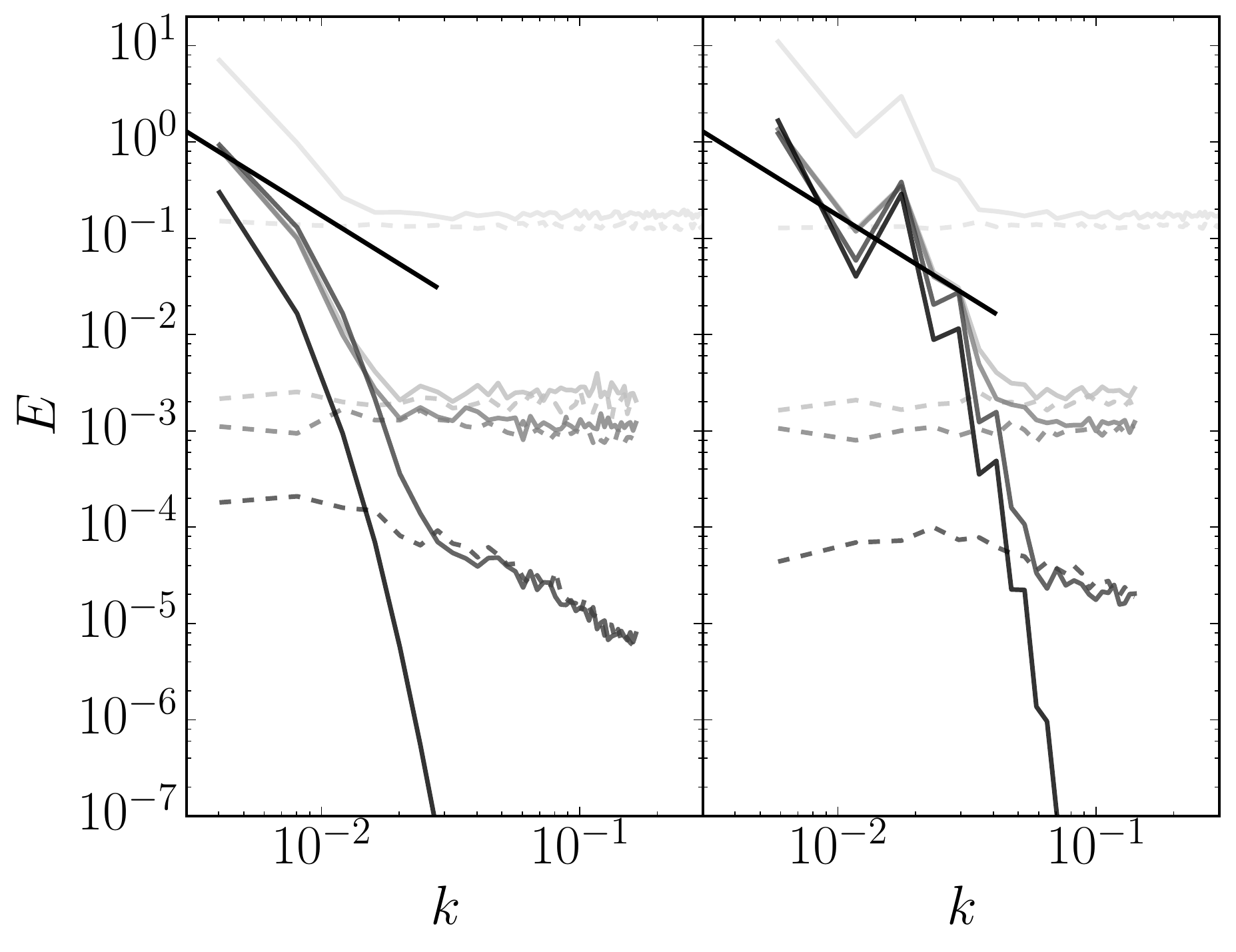}
\caption{The $x$ (left) and $z$ (right) axis spectral energy of the streamwise velocity $u$ at the channel centreline. 
\textit{a)} A fine grained single snapshot in turbulent (\textcolor{lllgrey}{$\xlinethick{}$}) and laminar (\textcolor{lllgrey}{$\xdashthick{}$}) flow.
\textit{b)} A single snapshot on a comparable grid to the CFD, turbulent (\textcolor{llgrey}{$\xlinethick{}$}) and 
laminar (\textcolor{llgrey}{$\xdashthick{}$}) flow. \textit{c)} An average of $64$ snapshots, turbulent (\textcolor{lgrey}{$\xlinethick{}$})
laminar (\textcolor{lgrey}{$\xdashthick{}$}) flow.  \textit{d)} $6400$ snapshots and three $y$ cell spatial average, turbulent (\textcolor{dlgrey}{$\xlinethick{}$}) and 
laminar (\textcolor{dlgrey}{$\xdashthick{}$}) flow. The CFD solution (\textcolor{ddlgrey}{$\xlinethick{}$}), with laminar case at limit of machine precision for all $k$.
Kolmogorov law  ($\xlinethick{}$) $E = C \epsilon k^{-5/3}$ for inertial subrange.}
\label{fft}
\end{figure}

In Figure \ref{fft}, four levels of grid refinement are considered in the MD channel compared to a single CFD case, as outlined in table \ref{table_fft}.
\begin{center}
\begin{tabular}{|c|c|c|}\hline
Case & $\tau_{_{MD}}$ &  Grid resolution \\\hline
\textit{a)} & $0.005$ & $672 \times 198 \times 400$ \\\hline
\textit{b)} & $0.005$ & $84 \times 198 \times 50$ \\\hline
\textit{c)} & $8$ & $84 \times 198 \times 50$ \\\hline
\textit{d)} & $32$ & $84 \times 66 \times 50$ \\\hline
\end{tabular}
\label{table_fft}
\end{center}
These are compared to a single CFD case with a $xz$ spectral resolution of $42 \times 42$ modes.
The Kolmogorov five thirds law is shown for reference, although due to the very low Reynolds number, no inertial range would be expected.
%
Each turbulent MD spectra is compared to the laminar MD solutions (dotted lines in Figure \ref{fft}) which have been averaged in the same manner.
These laminar solution are run at the same system size but with no turbulent initial condition.
The turbulent and laminar MD simulations show comparable base levels of broadband 'noise' in the spectra of Figure \ref{fft}, while only the turbulent case shows low wavelength structures.
These low wavelength structures exhibit similar shapes and magnitudes to the spectra obtained from the CFD solution.
One discrepancy is the lower energy in the CFD $x$ spectra relative to the molecular system. 
This may simply be a consequence of using different numbers of cells in the CFD and MD, the divergence of the trajectories or the higher effective Reynolds number in the MD simulation.
However, this may also be a consequence of the greater energy in an MD system due to the sub grid molecular fluctuations. 
This is supported by the results from case \textit{a)}, the finest grid resolution, where the entire spectral energy is much higher than observed in coarser grids. 
The level of thermal motion is at much greater energy and only two peaks are observed above this.
From Figure \ref{md_snapshotandzoom}, it is clear that this trend continues with increased grid refinement, until at the level of individual molecular velocities, even the streaks are difficult to identify.
The energy in the streamwise $u$ component, $\mathcal{K}_{flow}$, increases with a finer grid at the expense of the thermal energy, $ \mathcal{K}_{Them}$, as flow energy is split between thermal and streaming.
The higher spectral energy in the finest grid of Figure \ref{fft} suggests that coarse graining can result in lower energy in important flow structures.
In addition, the process of averaging results in spatial and temporal resolution being lost as the degrees of freedom are reduced.
However, without coarse graining, it would be impossible to identify the spectral content which may be of central importance to the regeneration cycle \citep{Hamilton}. 
For example in the $z$ spectra of Figure \ref{fft}, the roll structures give a large peak and a range of successive harmonics are clearly observed in the CFD solution. 
As the level of averaging is increased, spectral content previously obscured below the thermodynamic fluctuations becomes apparent in Fig \ref{fft}.
At least two further harmonics, previously hidden by the higher frequency thermal `noise' are uncovered.
The new peak are at the same wavelengths as a harmonic observed in the continuum solution.
Further low wavelegth peaks in the CFD spectra suggests that further harmonics would also become apparent in the MD spectra if even greater averaging was employed.
A range of flow length scales are clearly present in the MD minimal channel, as apparent in the energy spectra of Fig \ref{fft}, while the equivalent size of laminar channel does not display any hierarchy of scales.

One could question where the Kolmogorov microscale, $\eta = (\nu^3 / \epsilon)^{1/4}$, would be located on the spectrum of Figure \ref{fft}. 
Although the derivation of a Kolmogorov lengthscale assumes a much higher Reynolds number, consideration of the minimum eddy length scale at which coherent motion is dissipated to heat is a central concept in CFD modeling. 
Both for ensuring a DNS simulation is well resolved and identifying the minimum scale of turbulent eddy.
Although the minimum scale does not become apparent in the grid resolution study of Figure \ref{fft}, molecular dynamics may provide insight into the minimum scale of turbulent eddy.
If we assume that the Kolmogorov scale is below the observed noise floor and impossible to identify, we consider instead the minimum physical size this eddy could have in the molecular paradigm.
It is possible to view molecular liquids as a series of potential wells, created by the cage like configurational structure of a molecular fluid \citep{Zwanzig}. 
These molecular cages would typically see three dimensions motion of the molecule trapped within.
This provides an absolute minimum scale for a rotational vortex in a molecular simulation, displayed schematically on Figure \ref{cascade_schematic}.
Although these motions are likely different from the large scale vortical motions observed in the minimal channel, the interplay between inertial and the smallest scales of motions is of great interest.
Molecular dynamics is uniquely placed to provide insight into the minimum scale of eddy.

This section has demonstrated that the choice of grid has a major impact on the observed behavior of the system. 
In order to explore the velocity without requiring the definition of a grid, the probability density function are evaluated in the next section.
These distribution are compared to the probability density function of the cell averaged velocity.

\subsection{Probability Density Functions}
\label{sec:PDF}

\begin{figure}
        \centering
        \begin{subfigure}{0.45\textwidth}
                \includegraphics[width=\textwidth]{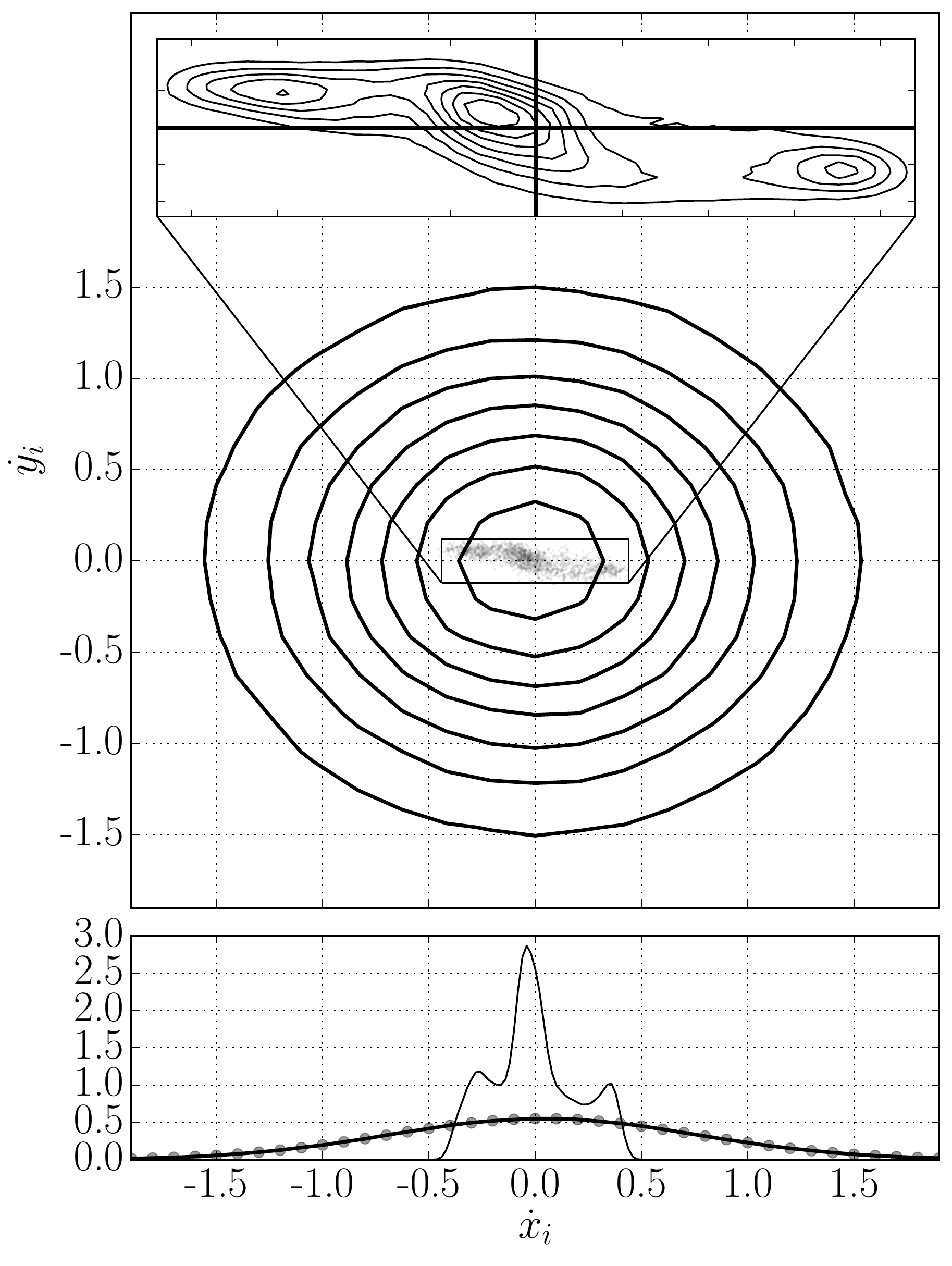}
                \caption{The joint probability density function for the molecular and averaged cell velocities. 
The bottom plot shows the $x$ projection of the molecular distribution  ($\xlinethick{}$) Gaussian fit ($\circ$) and CFD distribution ($\xline{}$). }
                \label{MD_PDF} 
        \end{subfigure} \;\;\; 
        \begin{subfigure}{0.45\textwidth} \;\;\;\;\;
                \includegraphics[width=\textwidth]{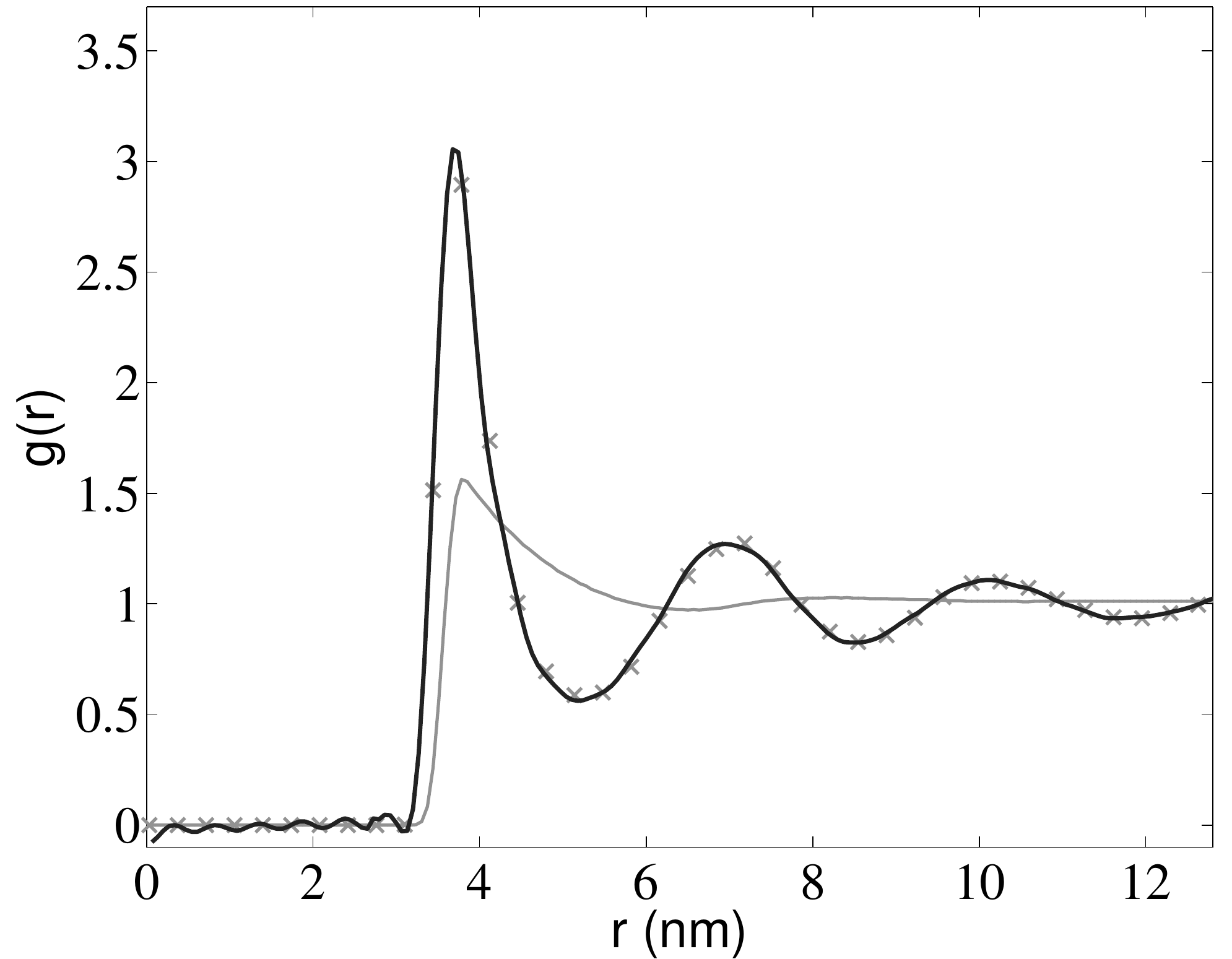}
                \caption{Radial distribution function for the current case $\rho=0.3$ and $T=0.4$ ($\textcolor{gray}{\xlinethick{}}$) with verification at $\rho=0.708$ and $T=0.8352$  ($\times$), equivalent to experimental results for liquid Argon at $T=85K$, $\rho = 0.02125 \text{\AA}^{-3}$ ($\xlinethick{}$).}
                \label{RDF}
        \end{subfigure}
        
        \caption{Probability density and Radial distribution functions}
        \label{PDF_RDF}
\end{figure}

The importance of cell based averaging techniques to the observation of flow structures was demonstrated in the previous subsection.
By averaging over time and space, information is essentially discarded about the range of fluctuations below the grid scale.
In this section, the distribution of molecular position and velocity are explored using probability density functions.
The joint probability density function (PDF) is displayed for individual molecular velocities and compared to velocity of the spatially averaged cells. 
The positional distribution of individual molecules is also considered in the form of the Radial distribution function.
The joint PDF is defined for each bin $n$ of width $\Delta u$ and height $\Delta v$,
\begin{align}
 P_{u,v}(u,v) = P([n-1]\Delta u < u < n\Delta u \text{ and } [n-1]\Delta v < v < n\Delta v),
\end{align}
with normalisation $\int_{-\infty}^\infty \int_{-\infty}^\infty  P_{u,v}(u,v) du dv = 1$.
The PDF for all the x and y velocities of the $\sim 1.4$ million individual molecules, $P_{u,v}(\dot{x}_i,\dot{y}_i)$ in a single $y$ cell at the channel centreline is shown in Figure \ref{MD_PDF}.
The distribution of the molecular velocities is almost perfectly Gaussian for both the $\dot{x}$ and $\dot{y}$ velocities.
This is demonstrated by the Gaussian fit shown on the  $\dot{x}$ projection at the bottom (similar results can be shown for $\dot{y}$).
Also shown in Figure \ref{MD_PDF} is the joint PDF of $P_{u,v}(u,v)$ where $u$ and $v$ are the time averaged ($\tau_{_{MD}}\!\! = 64$) velocity in the $84 \times 50$ cells at the channel centreline.
A total of $338$ $u$ and $v$ samples in time are used to give comparable statistics to the molecular case (a total of $\sim 1.4$ million samples).
These coarse grained cell velocities are shown as a scatter plot in Figure \ref{MD_PDF} with the zoom insert showing a contour plot of the PDF.
The $u$ projection of the CFD velocity PDF is also shown at the bottom of Figure \ref{MD_PDF}.
As there is negligible mean flow at the centreline, the average velocity in this top PDF are the perturbation velocities $u^\prime$ and $v^\prime$. 
The high speed $u$ streak clearly has an an associated positive $v$ velocity while the low speed has an associated negative $v$. 
The remaining distribution is distorted indicating the effect of the turbulent shear stretching.
There is a strong sweep and ejection behavior with a slight preference for ejection seen in the cell averaged PDF velocities \citep{Kevin_Nolan_2010}.
The sweep and ejection are though to be the main contribution to turbulent energy production and the cell averaged results obtained in Figure \ref{MD_PDF} are consistent with this observation.
The most striking conclusion, however, is that the probability density function of individual molecular velocities would simple obscure all of this detail.

The Radial Distribution Function (RDF) is shown in Figure \ref{RDF} and compared to neutron scattering experimental data for liquid Argon.
The RDF is calculated from \citep{Rahman_64} $g(r) = (V/N)[n(r)/4 \pi r^2 \Delta r]$ where $n(r)$ is the number of molecules in a spherical shell of size $\Delta r$.
To improve statistics, the RDF is calculated for every molecule in the system.
The available experimental data is at $\rho=0.708$ and $T=0.8352$ and the RDF for the equivalent state point in MD units is shown.
The RDF for the state point used in the turbulent study, $T=0.4$, $\rho=0.3$, is also included in figure \ref{RDF}.
It can be seen that the Lennard-Jones model, even for the short (WCA) cut off range used in this work, closely reproduces the structural properties of Argon.
This reproduction of the RDF also highlights the greatest strength of molecular dynamics; it is unique in being the only modeling technique to explicitly capture the underlying atomic structure.
The fluid's dynamics are governed by the interplay of molecular kinetic energy with the continual evolution of the current configuration of molecules.
As a result, the entire history of the molecular structure has a continual impact on the current state.
Pressure and viscosity are a consequence of the systems evolution and it is for this reason, the continuum style stress closure of \eq{STEqn} is not required in an MD model.

To explore the evolution of the impact of the molecules history on the flow, the trajectory of individual molecules are investigated in the next section.

\subsection{Lagrangian Statistics}
\label{sec:LS}

\begin{figure}
        \centering
        \begin{subfigure}{0.45\textwidth} \;\;\;\;\;
                \includegraphics[width=\textwidth]{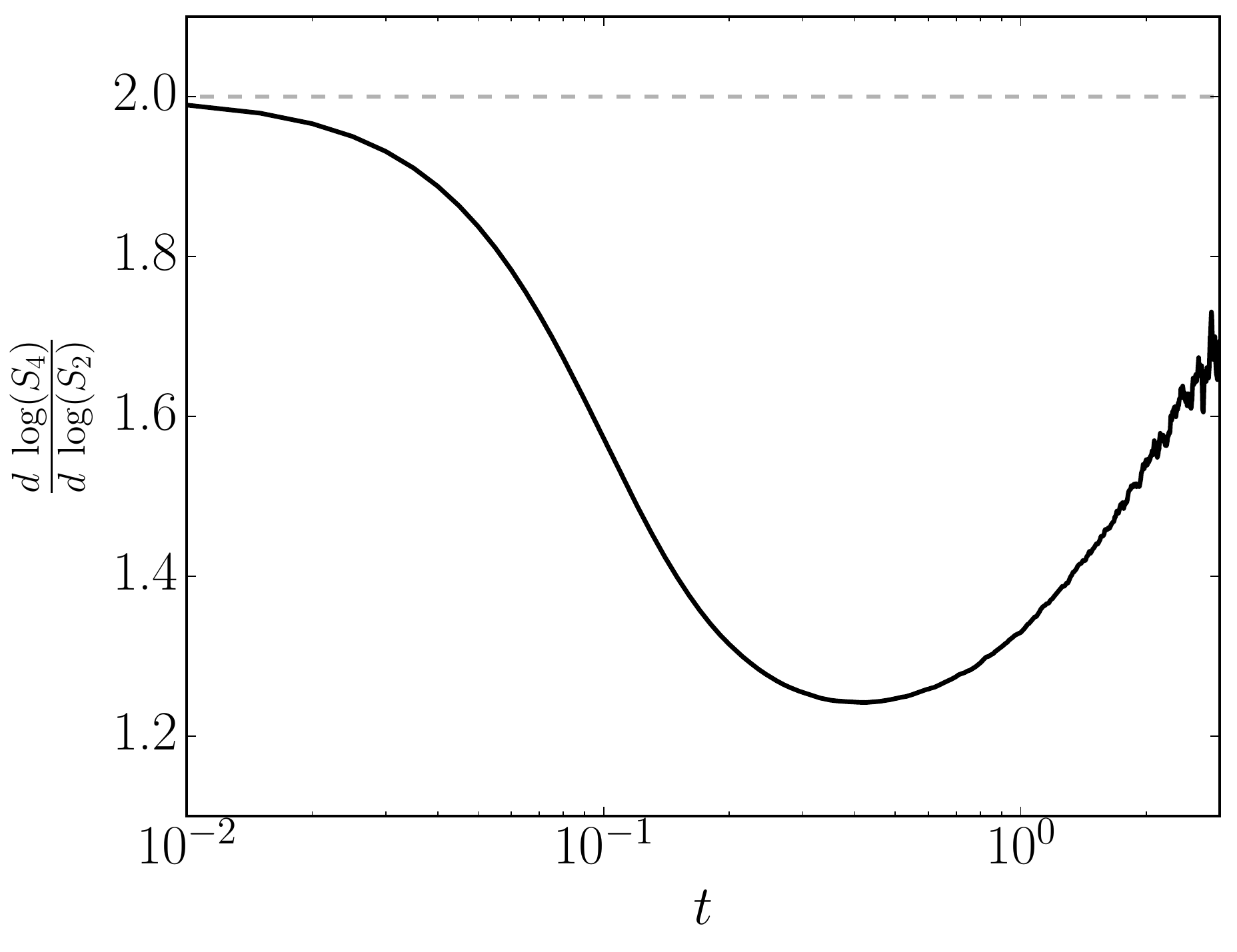}
                \caption{Logarithmic derivatives of structure functions \citep{Toschi_Bodenschatz}, }
                \label{structure_factor}
        \end{subfigure}
        \begin{subfigure}{0.45\textwidth}
                \includegraphics[width=\textwidth]{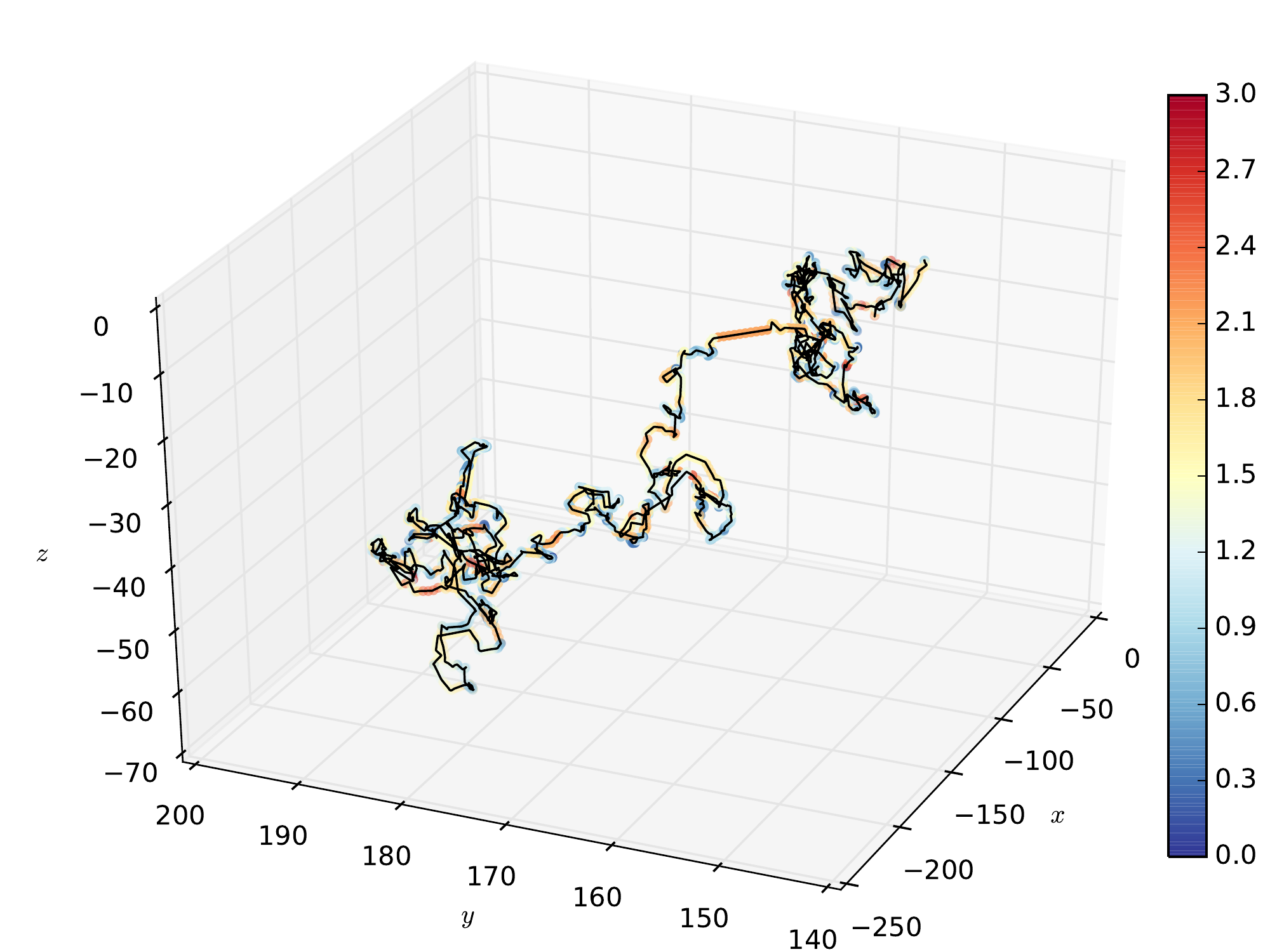}
                \caption{Trace of molecule coloured by magnitude of particle velocity.}
                \label{moltrace} 
        \end{subfigure} \;\;\; 
        \caption{Molecular trajectory and structure functions}
        \label{structure_factor_moltrace}
\end{figure}

\begin{figure}
        \centering
        \begin{subfigure}{0.45\textwidth} \;\;\;\;\;
                \includegraphics[width=\textwidth]{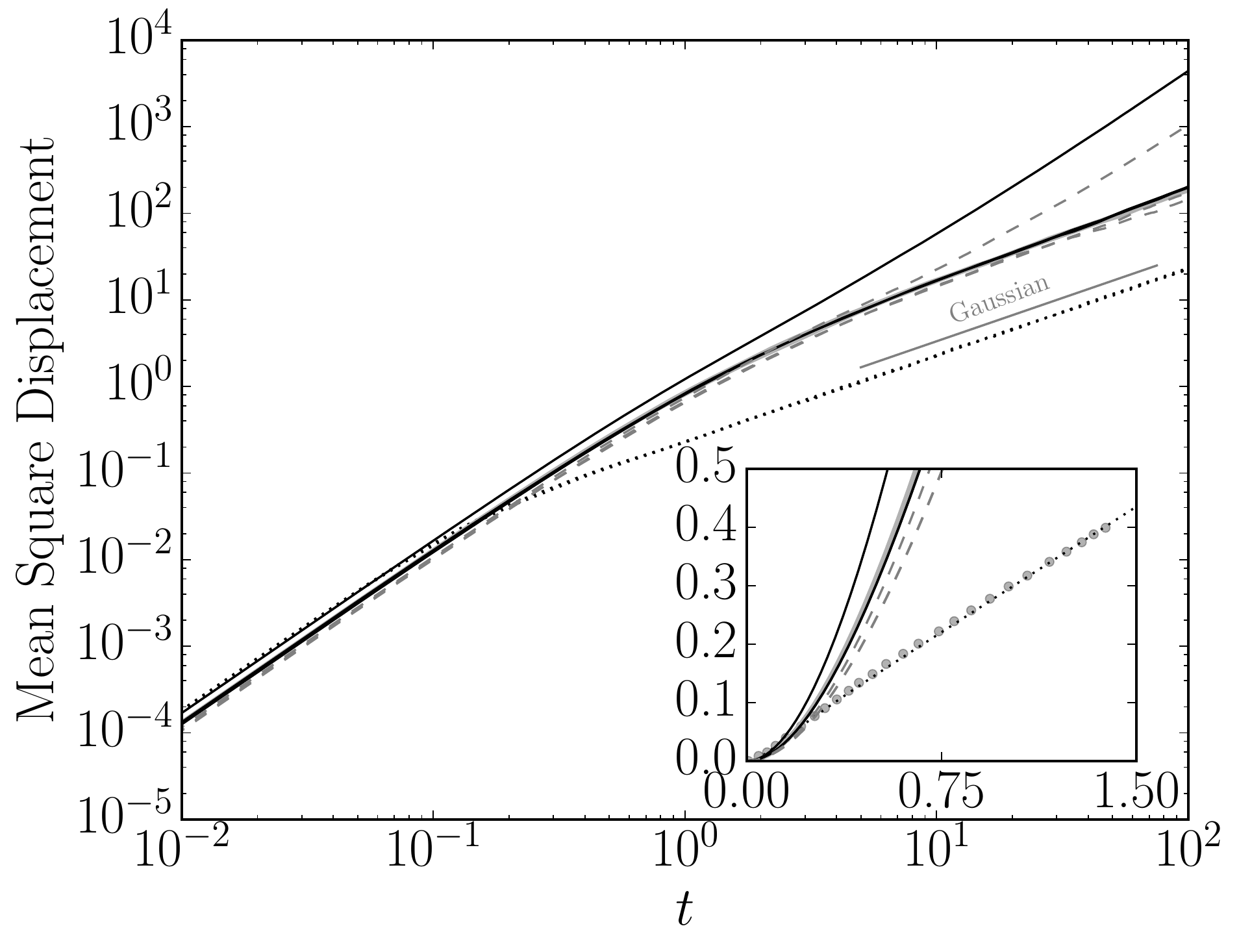}
                \caption{Mean square displacement, insert shows linear axis near zero and linear (Gaussian) line is shown and labelled.}
                \label{diffusion_fn}
        \end{subfigure}  \;\;\;\;\;
        \begin{subfigure}{0.45\textwidth}
                \includegraphics[width=\textwidth]{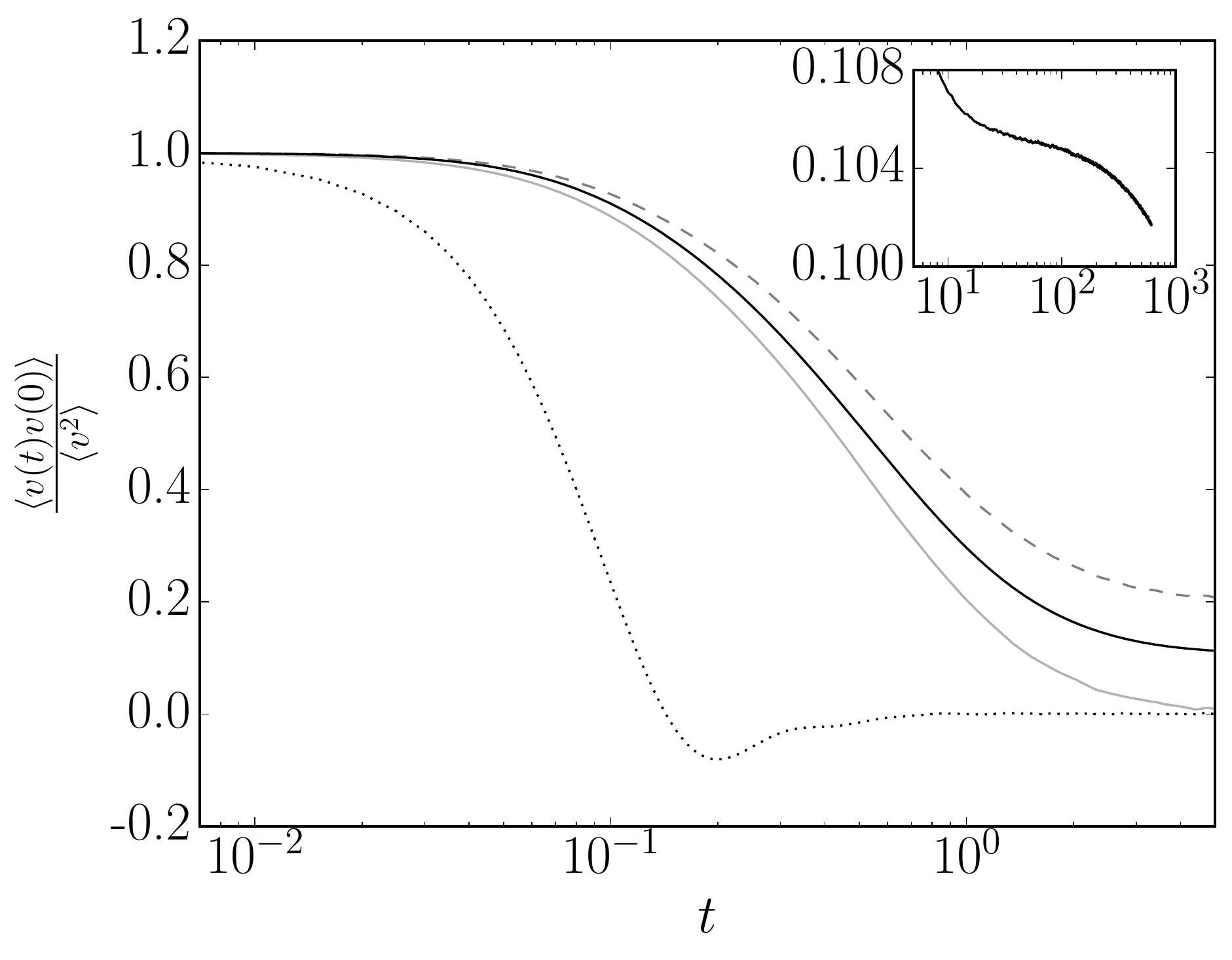}
                \caption{Autocorrelation of velocity, insert shows long time for minimal channel flow.}
                \label{autocorrel_fn} 
        \end{subfigure} \;\;\; 
        \caption{Molecular diffusion and autocorrelation functions. Results from \citet{Rahman_64} (\textcolor{gray}{$\bullet \!\!$ }) with matched LJ potential $r_c = 2.2$, $\rho=0.81$ and $T=0.78$ (\textcolor{black}{$\cdots$}), the WCA potential at $\rho=0.3$ and $T\approx0.5$ for an equilibrium case (\textcolor{lgrey}{$\xdashthick{}$}), wall driven Laminar flow (\textcolor{llgrey}{$\xlinethick{}$}) and the minimal channel flow (\textcolor{ddlgrey}{$\xlinethick{}$})}
        \label{diffusion_fn_autocorrel_fn}
\end{figure}

Lagrangian statistics have shown great promise as a means to understand the nature of turbulent flows \citep{Lagrange_stats_review}.
This involves experimentally following tracer particles, or in numerical studies, adding imaginary tracer particles into the turbulent flow.
In molecular dynamics the molecules themselves can be considered as tracer particles and Lagrangian statistics obtained from their evolution in time.
The distinction in molecular dynamics is that molecules are both tracers and the flow itself. 
A key parameter in Lagrangian statistics is the structure factor,
\begin{align}
S_p  = \frac{1}{N} \displaystyle\sum_{i=1}^N \left[ \boldsymbol{v}_i(t)  - \boldsymbol{v}_i(0) \right]^p  
\label{eq:structufact}
\end{align}
The logarithmic derivatives of the structure functions is used to probe intermittency \citep{ISI:000257283800027}.
The logarithmic derivatives of \eq{eq:structufact}, $S_4$ with respect to $S_2$, for all non-tethered molecules in the turbulent domain, is shown in Fig \ref{structure_factor}.
The resulting curve shows similar behaviour to studies of Lagrangian particles in turbulent flows \citep{ISI:000257283800027}.
The dip region in Fig \ref{structure_factor} has been associated with occurrence of small-scale vortex filaments in previous turbulence studies \citep{Toschi_Bodenschatz}. 
The presence of similar behaviour in a molecular simulation may be due to rotation inside molecular shells.
These rotating vortex like motions can be seen by observing the trajectory of a single molecule, as shown in figure \ref{moltrace}.
Much of the time is spent orbiting in molecular cages, although the meanflow in $x$ results in a much greater $x$ displacement over time.
Due to the low density of the system, the molecules will occasionally move large distances, L\'{e}vy flights, associated with super-diffusion \citep{Alder_Wainwright70}.
This behaviour can be seen by looking at the molecular mean squared displacement, defined as,
\begin{align}
\langle \boldsymbol{r}^2 \rangle = \frac{1}{N} \displaystyle\sum_{i=1}^N \left[ \hat{\boldsymbol{r}}_i(t)  - \boldsymbol{r}_i(0) \right]^2  
\end{align}
where the molecules positions are calculated with the $x$ meanflow removed $\hat{\boldsymbol{r}}_i(t+\Delta t) = \hat{\boldsymbol{r}}_i(t) + \Delta t \left[ v_i(t) - \overline{u}(t) \right]$.
The meanflow, $\bar{u}$, is obtained for a given molecules based on its wall normal position using a spline fit to the current velocity profile.
The mean square displacement is shown in Fig \ref{diffusion_fn}. 
The results from the turbulent flow are compared to a periodic equilibrium box and similar simulation of laminar flows with matched density $\rho=0.3$ and temperature $T \approx 0.5$.
Results for a dense fluid, $\rho=0.81$ and $T=0.78$, using the full LJ potential, $r_c = 2.25$, are also shown.
This is consistent with the work of \citet{Rahman_64} who verified the simulation diffusion using experimental data.
The behaviour is similar for all three systems, although despite removing mean velocity, the streamwise diffusion components remain larger.
This is attributed to the stick-slip behaviour near the walls which results in molecules being dragged at greater velocity than predicted from the mean profile.

In all cases, initially before the molecules collide, they move in the ballistic region with super-diffusive behaviour, $\langle \boldsymbol{r}^2 \rangle \propto t^\alpha$
Eventually, the mean squared displacement becomes linear as predicted by Gaussian statistics, $\langle \boldsymbol{r}^2 \rangle \propto t$, with the dense fluid case becoming linear earlier than the low density cases.
Perhaps surprisingly, both the laminar and turbulent flows observe the same mean square displacement as the equilibrium fluid.
It has been shown that diffusion is promoted by strong strain rates in a manner proportional to the square root of strain rate \citep{Cummings_et_al}.
As the domain is large and wall speed low, the strain rate is $ \gamma = 0.0035$ and the diffusion enhancement is minimal in both laminar and turbulent cases.
The vortices observed in the minimal channel seem to have negligible impact on diffusion, which appears to be dominated by the molecular structure for low shear rates.

The autocorrelation of velocity also appears to be dominated by the molecular structure, with a rapid decrease seen in the equilibrium, laminar and turbulent flow cases, Fig.\ \ref{autocorrel_fn}.
The differences in mean flow in the various cases explains the convergence of the autocorrelation to different final values for the various cases.
The dense fluid decorrelates more rapidly, exhibiting a negative region due to a bounce back after collisions.
Despite the dominance of molecular scale effects on short term autocorrelation, it appears that the hydrodynamics of the vortices may be evident at long times.
The insert in Fig.\ \ref{autocorrel_fn} appears to show a further decorrelation in velocity at longer times, possibility due to turbulent mixing
Confirmation of this behaviour would require longer studies to fully explore this phenomenon.
However, Lagrangian statistics applied to molecular dynamics shows great potential insight into the smallest scale of vortex dynamics in turbulent flow.


\section{Discussion}

In order to focus only on the fluid instability, in this work an incompressible and isothermal CFD solver was chosen for comparison. 
The molecular simulation includes compressibility, temperature and viscosity changes as well as thermal fluctuations. 
A more detailed continuum simulation could certainly include these features with compressibility and a changing viscosity coupled to the energy equation. 
In addition, a fluctuating hydrodynamics CFD solver could be employed to reproduce the thermal effects observed in the molecular channel.
The molecular dynamics minimal channel appears to show remarkably good agreement to the continuum model, despite compressibility and temperature change.
A lower temperature could be chosen or a more aggressive thermostatting mechanism could have been employed to remove this heat. 
However, the increase in temperature due to shear is a unique feature of molecular modeling and it is of interest to include it in the most realistic manner possible.
Despite not being in thermodynamics equilibrium, the breakdown and regeneration mechanism is robust enough to repeat twice in the presented work. 
The turbulent statistics, spectral range of energy scales and probability density function are in line with continuum observations. 
The presented study is clearly reproducing many of the features unique to turbulent simulation. 
By the end of the second regeneration cycle, the system energy appears to have reached a stable value, the Reynolds number if greater than $400$ and the minimal channel flow instability still appears to be present. 
The rate of increase in kinetic energy also appears to be almost zero by the end the second cycle.

The main feature of this flow, compared to previous NEMD studies, is that it includes time evolving hydrodynamics component as well as changing thermodynamics, resulting in a range of length and time scales.
The molecular definition of the kinetic pressure tensor, $\overline{\sum \langle \boldsymbol{p}_i 
\boldsymbol{p}_i/m_i  \rangle} $, and the Reynolds stress tensor, $\overline{\rho  \boldsymbol{u}^\prime \boldsymbol{u}^\prime
\vphantom{\boldsymbol{u}^\prime}}$, in \eq{Triple_decomp} are dimensionally and mathematically the same quantity on different timescales.
Osborne Reynolds actually proposed the decomposition of velocity, $u = \overline{u} + u^\prime$, by analogy with the kinetic theory \citep{Reynolds_1895}.
Fluctuations which contribute to Reynolds stress simply becomes kinetic pressure once the time and length scales of motion become small enough.
This is particularly interesting as certain researchers predicts a range of scales below the beginning of the dissipation range \citep{Frisch_Vergassola}.
The range of scales observed in the spectra of section \ref{sec:spectra}, 
the molecular and averaged probability density function \ref{sec:PDF} and vortex like motions in molecular trajectories of section \ref{sec:LS} 
all suggest that molecular studies may provide interesting insight into turbulence.

Reynolds shear stress and shear pressures play similar roles in turbulent and laminar flow respectively. 
Consider an MD simulation of steady state laminar Couette flow.
The shear stress from \eq{Stress} is given by the $x$ momentum flux and force on a plane in the $y$ direction,
\begin{align}
\Pi_{xy} dS_y = \frac{1}{2}  \bigg\langle  \displaystyle\sum_{i=1}^{N} 
 \displaystyle\sum_{i \ne j}^{N}  f_{xij}   dS_{yij} \bigg\rangle -  \bigg\langle \displaystyle\sum_{i=1}^{N} \frac{p_{xi} p_{yi}}{m_i }  dS_{yi}
\bigg\rangle = C_{laminar} 
\label{kinetic_config_balance}
\end{align}
On average the kinetic pressure and configurational stresses sum to a constant value in laminar flow.
The equality of \eq{kinetic_config_balance} is only true in a time averaged sense, with velocity and stresses fluctuating due to what is often termed molecular `noise'.
In turbulent Couette flow, a similar relation is seen by writing the Navier-Stokes equation in terms of Reynolds averaged quantities \citep{Aydin_Leutheusser},
\begin{align}
\overline{\Pi_{xy} dS_y}  - \rho \overline{u^\prime v^\prime} = C_{turbulent}
\label{Re_visc_balance}
\end{align}
The sum of Reynolds stress and viscous stress is a constant.
The form of Eqs. (\ref{kinetic_config_balance}) and (\ref{Re_visc_balance}) are strikingly similar, both representing a balance between shear effects due to fluctuations and a stress based shearing term.
Equation (\ref{Re_visc_balance}) is used to motivate the Boussinesq approximation \citep{potter_wiggert} in turbulent modeling which introduces an eddy viscosity coefficient to model the details of turbulent fluctuations.
Comparison with \eq{kinetic_config_balance} suggest that molecular detail is essential approximated by the continuum model in the form of a viscosity coefficient.
Due to a large number of molecules, $O(10^{25})$ molecules per $1m^3$ of air, the molecular viscosity is a statistically much better approximation than a turbulence based closure model.

%

One possible application of the present work is in the exploration of the transition to turbulence.
The effect of sub-continuum scales on turbulence is not known \citep{kim_et_al87} although it has been suggested that these microscale motions are important to the development of fluid instability \citep{Kadau_Review,Muriel}.
Some authors, even suggest the possibility that organized flow may actually originate from the smallest scales, rather than simply dissipate downward from the largest scales \citep{Tuck}.
Typically a CFD solution requires artificial random noise to trigger the transition from laminar to turbulent flow \citep{jimenez_moin}.
For an initially laminar case, the molecular level `noise' may provide a route to transition into a turbulent state as proposed, based on experimental evidence, by \citet{Muriel_2009}.
Indeed, recent work using fluctuating hydrodynamics have shown that thermal fluctuation provide a mechanism for the transition to turbulence \citep{de2013hydrodynamic}. 
To explore this, a low temperature MD simulation was performed with an effective Reynolds number of $Re \approx 700$ and initial velocity field based on steady state laminar Couette solution.
The flow remained laminar, suggesting that a larger system (and higher Reynolds number) would be required to observe transition from molecular fluctuations in this case.
As in a CFD simulation at $Re=460$, the MD solution remains laminar, suggesting that the inherent molecular noise, at least under the current conditions, is not sufficient to induce transition. 
In recent work, \citet{McWhirter} suggested that the global stability of MD Couette flow should be similar to its continuum counterpart. 
This could be further explored by introducing optimal perturbations to the flow or running the system at higher Reynolds numbers.

Another unique feature of molecular simulation is the ability to explore realistic surface fluid interactions.
In this study, the wall is solid argon, which forms a perfect face centered cubic crystal structure.
Wall-fluid interaction, $\epsilon_{wl}$, are set to unity, although more realistic walls, along with wall textures \citep{Jelly_et_al14} and polymer coatings, can be easily constructed in the molecular system.
Multiple fluid phases are trivially modelled by molecular systems which avoid the discontinuity presented by a three phase contact line and captures bubble nucleation.
The impact of rapid heating with phase change could be modelled for a turbulent channels, an important problem for coolant fluids in micro and nano scale devices.
A promising line of research applies molecular dynamics for fluid simulation as part of a coupled MD/CFD simulation.
In coupled simulation, an MD solver is employed in the near wall region and the bulk of the domain is efficiently modelled by CFD simulation \citep{OConnell_Thompson,Hadjiconstantinou_thesis,Mohamed_Mohamad,Smith_Thesis}.
As a results, molecular boundaries can be included in higher Reynolds number simulation and be used to understand the dynamic interplay of nano-scale boundaries and turbulent flow.
This is a logical extension of currently employed CFD coupling between near wall DNS in LES or RANS solvers to provide more detailed modeling at fluid solid interfaces.

This work has shown that the minimal flow unit, though to be a central component of near wall turbulence, can be modelled using a molecular dynamics description. The bursting cycle continues to occur in a similar manner to the incompressible continuum model, despite compressibility, significant heating and change of viscosity in the molecular model. This suggest that the non-dimensionalised Navier-Stokes equations model nonlinear effects which are potential valid to very small system sizes as shown by the good  agreement with the more fundamental molecular model.
The molecular model has great potential to provide insight into the nature of turbulent energy cascade below the diffusive range, the fluid-solid interaction and perhaps a mechanism for transition from the thermodynamics fluctuations which are an integral part of molecular simulation.


\section{Conclusion}

A minimal channel planar Couette flow has been simulated using molecular dynamics (MD).
The modeled MD fluid shows turbulent streaks breakdown and reformation with associated vortex regeneration consistent with experiments and computational fluid dynamics (CFD) simulations.
Reynolds averaged channel statistics show excellent agreement to both an equivalent CFD simulation and literature results.
A molecular form of the law of the wall exposes near wall molecular stacking due to stick slip behavior at the wall.
The occurrence of the regeneration cycle in MD, which is central to turbulent production, provides strong evidence that turbulence like behaviour can be reproduced in a molecular simulation.
The insight provided by molecular dynamics are explored through velocity spectra, probability density functions and Lagrangian statistics.
The similarity between the definition of molecular pressure and Reynolds stress suggests a continual cascade where eddying motions don't simply dissipate to incoherent heat at a certain scale.
In this way, molecular dynamics is a more fundamental model which requires no assumption of a viscosity coefficient and can explicitly models the full energy cascade.
As a result, simulation using molecular dynamics has great potential in the field of fluid dynamics turbulence research.
Simulating turbulence at the molecular scale can provide insight into the minimum turbulent eddy and the regeneration mechanism of turbulence from molecular origins; explore transition from laminar to turbulent flow; test the limitations of continuum models and aid the development of nano-scale modeling methodologies essential in many emerging technologies. 

\subsubsection{Acknowledgements}
I would like to thank D. Dini and D.M. Heyes, B. Todd,
P. Daivis, T.O. Jelly and D.J. Trevelyan for fruitful discussion and Simon Burbridge for his support in using the CX2 supercomputing resource at Imperial College. 
I would also like to thank J. Gibson for making the excellent Channelflow code open source.
This work is supported by an EPSRC Doctoral Prize Fellowship


\end{document}